\documentclass[letterpaper,twocolumn,10pt]{article}
\usepackage{usenix}

\usepackage{tikz}
\usepackage{amsmath}

\usepackage{filecontents}

\usepackage{amsmath,amssymb,amsfonts}
\usepackage{graphicx}
\usepackage{textcomp}
\usepackage{xcolor}
\def\BibTeX{{\rm B\kern-.05em{\sc i\kern-.025em b}\kern-.08em
    T\kern-.1667em\lower.7ex\hbox{E}\kern-.125emX}}

\usepackage[all]{nowidow}
\usepackage{placeins} 

\usepackage{listings}
\usepackage{url}
\usepackage{makecell}
\usepackage{hyperref}
\usepackage{adjustbox}
\usepackage{xspace}
\usepackage{multirow}
\usepackage{subfloat}
\usepackage{balance}
\usepackage{tikz}
\usepackage{ctable}
\usepackage{listings}
\usepackage[nobiblatex]{xurl}
\usepackage{mathrsfs}
\usepackage[normalem]{ulem}
\useunder{\uline}{\ul}{}

\usepackage{algpseudocode}
\usepackage{caption}

\usepackage{acronym}
\usepackage{fancybox}
\usepackage{verbatim}
\usepackage[normalem]{ulem}
\usepackage{float}
\usepackage{newfloat}
\usepackage{mdframed}

\usepackage{booktabs}

\usepackage{subfigure}
\usepackage[flushleft]{threeparttable}
\usepackage{booktabs}
\usepackage{tabularray}

\usepackage[utf8]{inputenc}
\usepackage[skins]{tcolorbox}
\usepackage{amsthm}

\usepackage{enumitem}
\usepackage{pifont}
\usepackage{tcolorbox}
\usepackage{subcaption}
\usepackage{multirow}
\usepackage[normalem]{ulem}
\usepackage{algorithm}
\usepackage{algpseudocode}

\usepackage{xspace}
\usepackage{ulem}

\definecolor{comp1}{RGB}{212,235,255}
\definecolor{comp2}{RGB}{255,245,200}
\definecolor{comp3}{RGB}{255,228,196}

\newcommand{\revise}[1]{{#1}}

\usepackage{amsthm}
\newtheorem{theorem}{Theorem}
\usepackage[dvipsnames]{xcolor}

\definecolor{RiseGreen}{RGB}{200,230,200}  
\definecolor{FallGray}{RGB}{230,230,230}   

\usepackage{siunitx}
\usepackage{graphicx}

\newcommand{\toolname}{\textsc{RAGCrawler}\xspace}
\newtheorem{definition}{Definition}

\definecolor{continuation}{RGB}{237, 125, 49}
\definecolor{keyword}{RGB}{68, 114, 196}
\definecolor{ours}{RGB}{84,130,53}

\usepackage{fontawesome5}

\newcommand{\authicon}{\textcolor{black!60}{\faIcon{user-lock}}\xspace}         
\newcommand{\repicon}{\textcolor{black!60}{\faIcon{sync-alt}}\xspace}     
\newcommand{\diskicon}{\textcolor{black!60}{\faIcon{save}}\xspace}      
\newcommand{\depicon}{\textcolor{black!60}{\faIcon{cube}}\xspace}       

\begin{document}

\date{}

\title{Connect the Dots: Knowledge Graph–Guided Crawler Attack on Retrieval-Augmented Generation Systems}

\author{
{\rm Mengyu Yao\textsuperscript{1}, Ziqi Zhang\textsuperscript{2}, Ning Luo\textsuperscript{3}, Shaofei Li\textsuperscript{1}, Yifeng Cai\textsuperscript{1}, Xiangqun Chen\textsuperscript{1}, Yao Guo\textsuperscript{1} and Ding Li\textsuperscript{1}}\\
\textsuperscript{1}MOE Key Lab of HCST (PKU), School of Computer Science, Peking University\\
\textsuperscript{2}Department of Computer Science, University of Illinois Urbana-Champaign\\
\textsuperscript{3}Department of Electrical and Computer Engineering, University of Illinois Urbana-Champaign\\
}




\maketitle

\begin{abstract}
Stealing attacks pose a persistent threat to the intellectual property of deployed machine-learning systems. Retrieval-augmented generation (RAG) intensifies this risk by extending the attack surface beyond model weights to knowledge base that often contains IP-bearing assets such as proprietary runbooks, curated domain collections, or licensed documents.  
Recent work shows that multi-turn questioning can gradually steal corpus content from RAG systems, yet existing attacks are largely heuristic and often plateau early.
We address this gap by formulating {RAG knowledge-base stealing} as an {adaptive stochastic coverage problem} (ASCP), where each query is a stochastic action and the attacker’s goal is to maximize the {conditional expected marginal gain} (CMG) in corpus coverage under a query budget. 
Bridging ASCP to real-world black-box RAG knowledge-base stealing raises three challenges: CMG is unobservable, the natural-language action space is intractably large, and feasibility constraints require stealthy queries that remain effective under diverse architectures. 
We introduce \toolname, a knowledge graph-guided attacker that maintains a global attacker-side state to estimate coverage gains, schedule high-value semantic anchors, and generate non-redundant natural queries.
Across four corpora and four generators with BGE retriever, \toolname achieves 66.8\% average coverage (up to 84.4\%) within 1,000 queries, improving coverage by 44.90\% relative to the strongest baseline. It also reduces the queries needed to reach 70\% coverage by at least $4.03\times$ on average and enables surrogate reconstruction with answer similarity up to 0.699. Our attack is also scalable to retriever switching and newer RAG techniques like query rewriting and multi-query retrieval. 
These results highlight urgent needs to protect RAG knowledge assets.
\end{abstract}

\section{Introduction}\label{sec:intro}



Intellectual property (IP) theft remains a central concern across the AI service landscape.
A large body of work on model-stealing attacks and defenses~\cite{juuti2019prada,oliynyk2023know,carlini2024stealing,rakin2022deepsteal,orekondy2019knockoff,shen2022model,tramer2016stealing,he2021stealing,liu2022stolenencoder,yuan2022attack,das2025siamese,danda2021towards,orekondy2020prediction,li2022defending,kariyappa2020defending,wu2024efficient} underscores the urgency of protecting proprietary AI assets and preventing information theft.
As modern AI service architectures grow increasingly complex, the IP attack surface extends beyond the model itself, requiring immediate investigation and comprehensive defense strategies.

Retrieval-Augmented Generation (RAG), now a ubiquitous architecture grounded in proprietary data, extends the attack surface to the retrieval corpus that sits on the path of every response.
RAG couples a generator with a retriever over a document collection, using retrieved documents to improve factuality and domain adaptation~\cite{karpukhin2020dense,lewis2020retrieval}.
In enterprise and vertical deployments, the corpus often contains IP-bearing assets (e.g., proprietary runbooks and policies, curated domain data, and licensed materials) whose acquisition and maintenance are costly~\cite{xu2024generative,al2023transforming,loukas2023making}.
\revise{Because each response is grounded in retrieved documents, the public chat or API interface~\cite{vertexaisearch_pricing} becomes an attack surface: an attacker can progressively steal more of the underlying corpus using curated queries.
}
Stealing a sufficiently large portion of the corpus can be enough to replicate the service: an attacker can pair the stolen corpus with an off-the-shelf LLM to build a substitute RAG assistant that approximates the victim’s behavior without access to the original deployment~\cite{wang2025silent}.

Recent studies highlight that \emph{knowledge-base stealing} poses an important and growing threat to deployed RAG systems, with direct copyright, licensing, and privacy implications for IP-bearing corpora~\cite{zeng2024good,qifollow,jiang2024rag,wang2025silent}.
Attackers can elicit verbatim or near-verbatim spans from RAG outputs~\cite{zeng2024good,qifollow}, and sustained multi-turn interaction can cumulatively expand the portion of a hidden corpus that becomes exposed~\cite{wang2025silent,jiang2024rag}.
However, existing attacks typically generate prompts by locally reacting to the most recent response, either continuing the thread or pivoting on extracted keywords~\cite{jiang2024rag,wang2025silent}.
Such heuristics often drift off-topic (left panel of Fig.~\ref{fig:example}~(a)) or revisit already-exposed regions (middle panel of Fig.~\ref{fig:example}~(a)), wasting query budget and plateauing early, leaving much of the corpus unexplored (Fig.~\ref{fig:example}~(b)). Moreover, while prior empirical studies demonstrate the knowledge-base theft, theoretical grounding and provable guarantees for coverage-maximizing attacks remain unknown.
This motivates a critical question: \emph{can the adversary steal the data more efficiently, and where is the limit?}

To address this question, we identify the fundamental gap in existing work: \emph{lack of a principled objective to maximize {global} corpus coverage}. Without such an objective, query selection is driven by local reactions to the latest response, providing no globally consistent criterion to avoid redundancy and systematically expand coverage across turns. An effective knowledge-base stealing strategy should instead exploit the \emph{global state} accumulated over the entire interaction history and choose each query by its \emph{expected marginal contribution} to previously unseen corpus content.


We formalize this global objective by casting RAG knowledge-base stealing as \emph{RAG Crawling}, an instance of the \textbf{Adaptive Stochastic Coverage Problem (ASCP)}~\cite{nemhauser1978analysis,wolsey1982analysis,golovin2011adaptive}. 
In our reduction, each query is an action that stochastically reveals a subset of hidden documents through retrieval, and the attacker aims to maximize the expected coverage of unique corpus items under a fixed query budget. 
ASCP provides an adaptive-greedy benchmark that selects the query with the largest \emph{conditional expected marginal gain (CMG)} at each step. 
We further prove that \emph{RAG Crawling} satisfies adaptive monotonicity and adaptive submodularity; thus, under the same budget of $B$ queries, the CMG-based adaptive greedy attack is provably near-optimal, achieving a $(1-1/e)$-approximation to the expected corpus coverage attainable by an optimal adaptive attacker.



However, implementing this theory into a practical black-box attack faces three key instantiation challenges. 
%
\ding{182}~\textbf{CMG is unobservable:} the attacker cannot directly measure the coverage increment induced by a query and its retrieval outcome without corpus access, so the true CMG cannot be exactly computed.
\ding{183}~\textbf{The action space is intractable:} the query space $Q$ is effectively unbounded (any natural-language string), making it infeasible to exhaustively search for the CMG-maximizing query without additional structure.
\ding{184}~\textbf{Real-world feasibility constraints are strict:} queries must remain natural and innocuous to avoid detection or refusal, and high-gain intents must be expressed in policy-compliant surface forms that remain stable under query rewriting and multi-query retrieval.

To solve these challenges,
we introduce \toolname, a novel attack framework consisting of three stages: knowledge graph construction, strategy scheduling, and query generation, that together overcome the above challenges.
\toolname maintains an attacker-side knowledge graph (KG) that provides a global summary of acquired information and supports planning. 
With this global state, \toolname first uses KG growth in KG construction phase to estimate coverage gain and CMG.
\toolname next prunes the action space by selecting high-value semantic anchors in strategy scheduling phase. 
These semantic anchors are determined based on historical gains and the structural exploration of the KG.
During the query generation phase, \toolname enforces real-world feasibility by turning top-ranked anchors into fluent, policy-compliant natural-language queries with history-aware deduplication.
Together, these three stages form a practical instantiation of the adaptive greedy policy for ASCP.

Experiments across diverse deployments show that \toolname substantially improves RAG knowledge-base stealing effectiveness. Across four datasets and four RAG generators (including safeguard variants), it achieves up to 84.4\% corpus coverage within a 1{,}000-query budget with BGE retriever. It improves coverage by 44.90\% relative to the best baseline and reduces the queries needed to reach 70\% coverage by at least 4.03× on average. The stolen content also supports service replication: a surrogate RAG built from the recovered corpus reaches up to 0.699 answer similarity to the victim system. \toolname is scalable to other RAG techniques like retriever switching, query rewriting, and multi-query retrieval.

\noindent{\bf Our contribution.} 
\begin{itemize}[topsep=0pt, leftmargin=1em, noitemsep]
    \item \textbf{Theory.} We formalize {RAG knowledge-base stealing} as {RAG Crawling}, an adaptive stochastic coverage problem. We provide a principled, objective, and theoretical grounding for coverage-maximizing black-box attacks.
    \item \textbf{System.} We design \toolname, a knowledge graph--guided crawler that instantiates the adaptive greedy coverage principle under real-world constraints.
    \item \textbf{Evaluation.} We conduct extensive evaluations across diverse settings to demonstrate the effectiveness of \toolname, as well as against modern RAG defenses.
\end{itemize}

\section{Motivating Example}
\label{sec:motivating-example}

\begin{figure*}[t]
    \centering
    \includegraphics[width=0.77\linewidth]{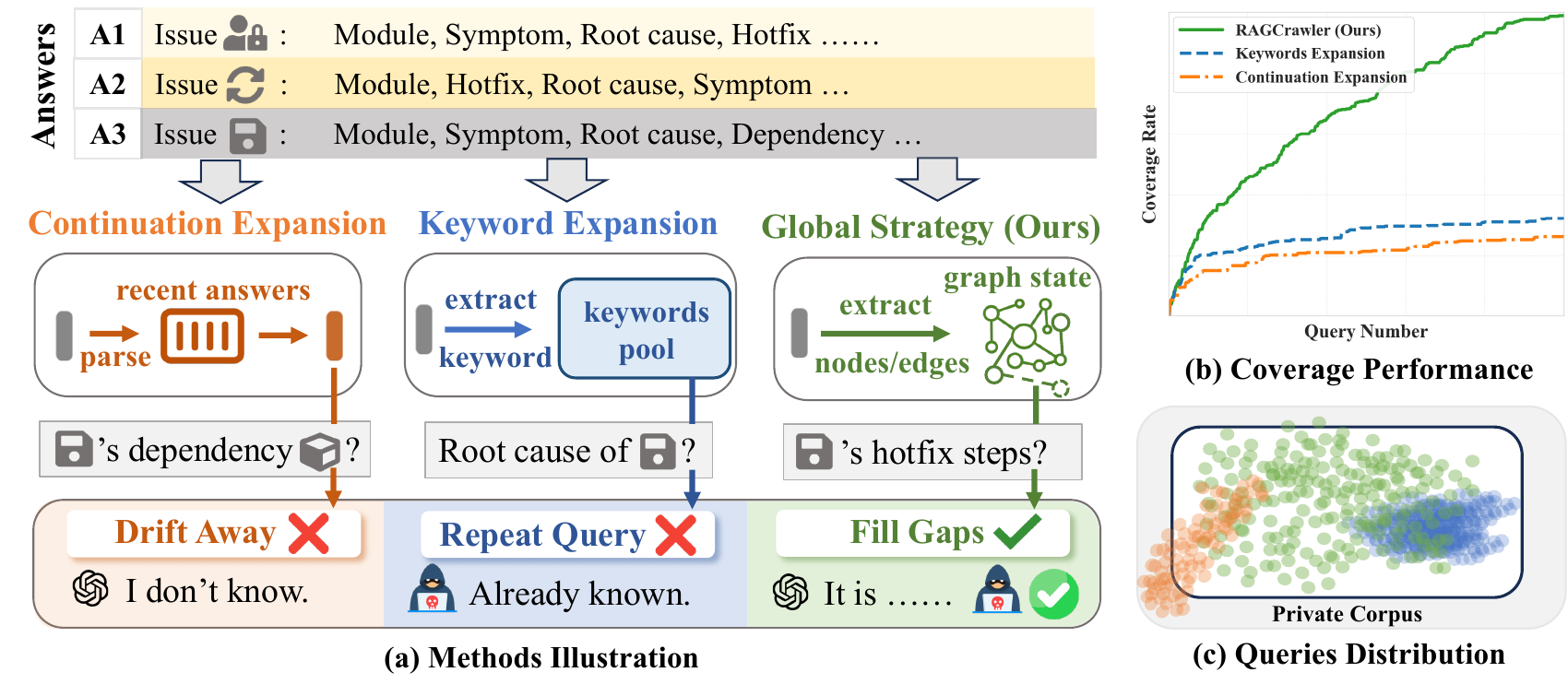}
    \caption{\textbf{A motivating example.}
    (a) A {RAG knowledge-base stealing} scenario on an IP-bearing troubleshooting corpus. Upper answers represent knowledge obtained from earlier attack rounds.
    The continuation strategy follows the recent answer, which can drift away from the corpus;
    the keyword strategy pivots on extracted keywords, often yielding redundant information from the same semantic region;
    our proposed strategy maintains a global graph, detects unrevealed facts and targets them with new queries.
    (b) Knowledge coverage w.r.t query number for each strategy on an example corpus.
    (c) Distribution of retrieved document in the corpus’ semantic space. Each point is a document, with colors indicating queries from different strategies.}
    \label{fig:example}
\end{figure*}

Consider an attacker targeting a paid support troubleshooting RAG assistant, which is backed by a vendor support portal (Fig.~\ref{fig:example}(a)).
Such portals are often built on private issue tracking tickets and internal resolution notes, and are usually considered IP assets~\cite{jira}. 
In our example, we will select three representative issues and assign an icon to each.
\diskicon (storage), \repicon (replication) , and \authicon (authentication) represent three issues.
The attacker’s goal is to steal as much of the hidden knowledge base as possible through multi-round queries. 
By repackaging the stolen knowledge, attackers can mount further downstream attacks such as rehosting the RAG system, bypassing access controls, undercutting subscriptions, or violating licensing and trade secret protections~\cite{wang2025silent}.

Existing stealing attacks mainly rely on two local heuristics, \textcolor{continuation}{\bf continuation expansion} and \textcolor{keyword}{\bf keyword expansion}.
In the first strategy (e.g., RAGThief~\cite{jiang2024rag}), the new query is built on the context of existing answers.
The problem of this strategy is that the query can easily \textcolor{continuation}{\bf drift away} from the actual corpus content.
For example, in Fig.~\ref{fig:example}~(a), the last answer about Issue~\diskicon\ attributes the failure to a dependency package \depicon.
A continuation-based attacker will ask about \depicon\ next, but since it is not in the licensed knowledge base’s scope, this line of questioning yields no useful information.
A \textcolor{keyword}{\bf keyword-based strategy} (e.g., IKEA~\cite{wang2025silent}) pivots each new query around key terms of previous answers, keeping it mostly within the same semantic neighborhood of the corpus.
For example, in Fig.~\ref{fig:example}~(a), the keyword-based attack \textcolor{keyword}{\bf repeatedly asks} about \diskicon's root cause and does not explore new contents.

\begin{figure}[b!]
    \centering
   
    \includegraphics[width=0.7\linewidth]{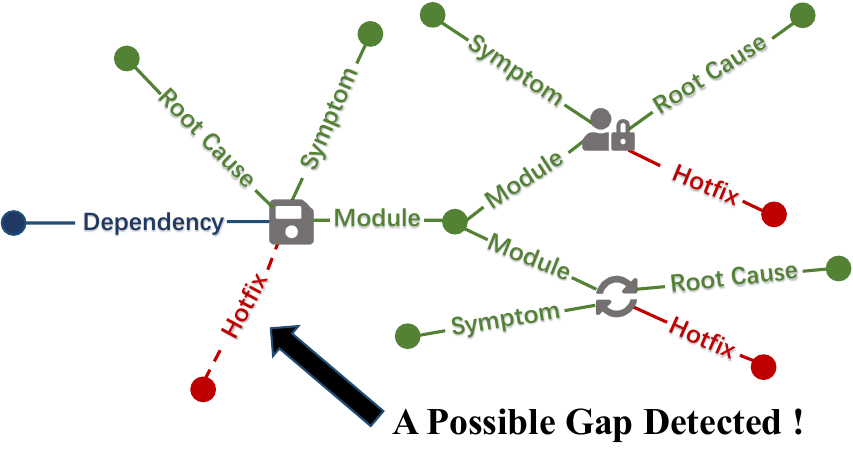}
    \caption{\textbf{Illustration of gap detection by the Global Strategy (Ours).}
    The strategy identifies from the graph that the storage issue (\diskicon) has a missing \emph{Hotfix} facet, and guides the next exploration step to complete this unrevealed knowledge.}
    \label{fig:concrete-graph}
\end{figure}

Our approach maintains a \textcolor{ours}{\bf growing knowledge graph} that summarizes discovered entities and relations for globally-planned exploration. Fig.~\ref{fig:concrete-graph} shows the knowledge graph of our attack. Nodes represent entities (e.g., issues, hotfixes, modules), and edges capture relations or co-occurrences (e.g., issue $\rightarrow$ hotfix).
As new answers arrive, the graph is expanded with all new entities or relations.
By maintaining this global knowledge state, the attacker can detect gaps in what has been revealed so far.
As illustrated in Fig.~\ref{fig:concrete-graph}, the knowledge graph dynamically reveals incomplete regions of information.
For instance, the graph shows that the issue \diskicon\ has no revealed \emph{Hotfix}, while the other issues
(\authicon\ and \repicon) already have their hotfix information identified.
Then the next query will be directed to the uncovered data of \diskicon’s unobserved hotfix steps, expanding overall coverage.

Our empirical results show that using this global knowledge graph significantly improves the coverage rate.
Fig.~\ref{fig:example}~(b) shows the knowledge coverage rate vs.\ the number of queries for our attack and two baseline strategies.
Guided by the global knowledge graph, our attack (green line) converges much faster than continuation-based
(orange curve; focusing on unrelated queries) and keyword-based (blue curve; focusing on repetitive queries) strategies.
Fig.~\ref{fig:example}~(c) provides a qualitative view of how each strategy explores the semantic space of the corpus. Each point denotes a document, and the black box indicates the knowledge-base space. Guided by the global knowledge graph, our attack (green points) spread widely across the space and covers many different clusters of information. The continuation-based queries (orange points) follow a narrow path through the space, gradually drifting away from the space center. Keyword-based queries (blue points) form a few dense clusters around specific topics.



\section{Background}

\subsection{Retrieval-Augmented Generation Systems}
\label{sec:bg-rag}
A Retrieval-Augmented Generation (RAG) system $\mathcal{S}$ typically comprises a generator $\mathcal{G}$ (an LLM), a retriever $\mathcal{R}$, and a document collection $\mathcal{D}$. Given a user query $q$, the retriever $\mathcal{R}$ begins by producing an embedding for $q$ and based on some similarity function (typically cosine similarity), fetching the k most relevant documents:
\begin{equation}\label{equation:rag-retrieve}
    \mathcal{D}_k(q) = \arg\text{ top-}k_{d \in \mathcal{D}} \, \text{sim}(q, d).
\end{equation}
where $sim(\cdot)$ represents the similarity function, and $\arg\text{ top-k}$ selects the top-k documents with the highest similarity scores. The generator $\mathcal{G}$ then produces an answer conditioned on the query and retrieved context \cite{lewis2020retrieval}:
\begin{equation}\label{equation:rag-gen}
    a = \mathcal{G}\big(\text{wrapper}(q, \mathcal{D}_k(q))\big),
\end{equation}
where $\text{wrapper}(\cdot,\cdot)$ denotes the system prompt that interleaves $q$ with the retrieved documents. 

In practice, deployed RAG pipelines often incorporate guardrails or enhancements. The most commonly used are \emph{query rewriting} and \emph{multi-query retriever}.

\noindent{\bf Query rewriting}~\cite{lin2020conversational,ma2023query,mo2023convgqr,wang2025maferw} transforms the original query to improve retrievability, resolve ambiguity, repair typos, or strip unsafe instructions. The downstream retrieval and generation processes then operate on the rewritten query. 

\noindent{\bf Multi-query retrieval}~\cite{multiquery,li2024dmqr} generates multiple paraphrases of the original query and retrieves independently for each, aggregating the results into a candidate pool. Rather than concatenating all retrieved documents, RAG systems typically re-rank candidates to select the most relevant ones, with \emph{Reciprocal Rank Fusion (RRF)}~\cite{cormack2009reciprocal} widely used. While not a dedicated defense, multi-query retrieval’s altered retrieval dynamics may change the attack surface.

\subsection{Adaptive Stochastic Coverage Problem}
\label{sec:bg-ac}


The \emph{Adaptive Stochastic Coverage Problem} (ASCP) models budgeted coverage maximization in sequential decision-making under uncertainty. At each step, an agent selects an action, observes what it reveals, and adapts future actions based on the observations so far. 
The objective is to maximize expected coverage under a limited budget, such as a limited number of actions.
An ASCP instance can be specified by: 

\begin{itemize}[leftmargin=10pt,noitemsep,topsep=0pt]
  \item \textbf{Item universe ($\mathcal{U}$)}: The item universe $\mathcal{U}$ represents the set of all items that are available for coverage.

  \item \textbf{Action space ($\mathcal{Q}$)}: The action space $\mathcal{Q}$ consists of all possible actions. Each action $q$ can potentially reveal (cover) a subset of items in $\mathcal{U}$.

  \item \textbf{Stochastic outcome (realization $\Phi$ and $O_{\Phi}(q)$)}: Uncertainty is modeled by a latent realization $\Phi$ (a ``state of the world'') drawn from a distribution. Conditioned on $\Phi$, executing action $q$ reveals an outcome set $O_{\Phi}(q)\subseteq \mathcal{U}$.

  \item \textbf{Cost and budget ($c(q), B$)}: The cost $c(q)$ represents the resources required to perform action $q$, and $B$ is the total available budget. 

  \item \textbf{Coverage function ($f(S,\Phi)$)}: The coverage function $f(S,\Phi)$ quantifies the utility of what has been revealed by a set of actions $S$ under realization $\Phi$. For coverage objectives, $f$ is commonly a function of the union $\bigcup_{q\in S} O_{\Phi}(q)$.
\end{itemize}

\begin{theorem}[Approximation guarantee of adaptive greedy~\cite{nemhauser1978analysis,wolsey1982analysis,krause2007near,golovin2011adaptive,khuller1999budgeted}]
\label{theorem:adaptive-greedy}
Assume $f$ is adaptively monotone and adaptively submodular. Under a cardinality budget $B$, the adaptive greedy policy $\pi_{\mathrm{greedy}}$ (maximizing CMG at each step) achieves a $(1-1/e)$-approximation to the optimal expected coverage\footnote{$(1-1/e)\!\approx\!0.63$} attained by the optimal adaptive policy $\pi^\star$ under same budget.
\end{theorem}



We reduce \emph{RAG knowledge-base stealing} to ASCP by formalizing it as \emph{RAG Crawling} (Sec.~\ref{sec:problem-formulation}), enabling CMG-based greedy optimization as a principled strategy for globally planned stealing under a query budget.

\subsection{Knowledge Graph}
\label{sec:bg-kg}
Knowledge graph (KG) is a widely-used knowledge representation technology to organize huge amounts of scattered data into structured knowledge. Formally, a KG is a directed, labeled graph $\mathcal{G}=(\mathcal{E},\mathcal{L},\mathcal{R})$, where $\mathcal{E}$ is the set of entities, $\mathcal{R}$ is the set of relation types (edge labels), and $\mathcal{L}\subseteq \mathcal{E}\times\mathcal{R}\times\mathcal{E}$ is the set of labeled, directed edges (facts). Each edge $l\in\mathcal{L}$ is a triple $(h,r,t)$ that connects head $h\in\mathcal{E}$ to tail $t\in\mathcal{E}$ via relation $r\in\mathcal{R}$. For example, $(\text{Python},\ \texttt{instanceOf},\ \text{Programming Language})$.

Compared with unstructured text, KGs provide schema-aware, canonicalized facts with explicit connectivity and constraints~\cite{ji2021survey}. We leverage the knowledge graph to construct an attacker-side state that makes coverage estimable, constrains exploration to a compact semantic action space, and records provenance for robust, auditable updates.

\section{Threat Model}
\label{subsec:threat-model}

\begin{figure}[tp]
    \centering
    \includegraphics[width=0.75\linewidth]{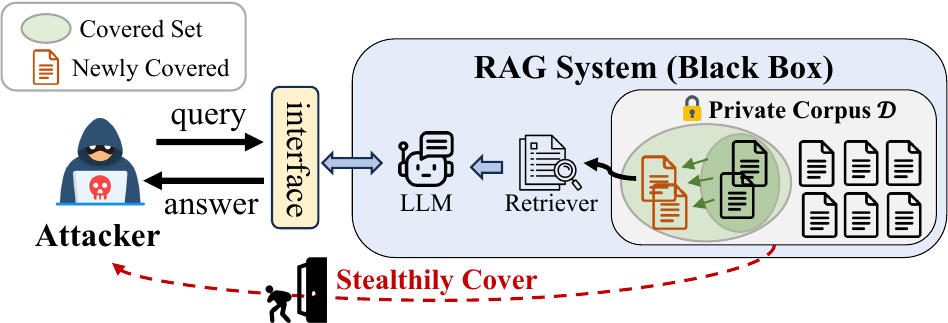}
    \caption{\textbf{Threat model.}
    An external attacker has \emph{black-box} access to a deployed RAG system and can only interact via its public query interface. 
    For each query, the retriever selects a set of documents from the corpus and the LLM generates an answer conditioned on these documents.
    Across turns, the attacker aims to stealthily expand the union of leaked retrieved items under a query budget.}
    \label{fig:threat-model}
\end{figure}

\noindent\textbf{Scenario.}
As illustrated in Fig.~\ref{fig:threat-model}, we consider a production RAG system accessed through a black box interface such as an API or chat. Internally, a retriever selects passages or documents from a corpus $\mathcal{D}$ and a generator composes an answer from the query and the retrieved documents~\cite{karpukhin2020dense,lewis2020retrieval}. In deployment the generator may rewrite, summarize, or refuse according to guardrails and access policies~\cite{multiquery,beck2025ragqr}. The service does not reveal retrieval scores or document identifiers. The attacker has no insider access and interacts as a normal user. The attacker aims to steal as much of $\mathcal{D}$ as possible  by steering retrieval to cover new items across turns.

Prior work has identified two classes of such stealing attacks in RAG systems: \emph{Explicit} attacks append overt adversarial instructions to elicit verbatim disclosure of retrieved text~\cite{cohen2024unleashing,jiang2024rag,di2024pirates,zeng2024good}.
\emph{Implicit} attacks use benign-looking queries and aggregate factual content that the system legitimately returns across turns, reconstructing corpus via paraphrases and fragments~\cite{wang2025silent}.
Our focus is the implicit regime, which better matches real usage and evades refusal rules.

\noindent\textbf{Adversary's Goal.}
The attacker aims to steal the knowledge base $\mathcal{D}$ as completely as possible while remaining stealthy.
Leakage may appear as verbatim text, paraphrases, summaries, or factual statements grounded in retrieved documents.
At round $t$, the adversary submits $q_t$ and the system internally consults a hidden retrieved set $\mathcal{D}_k(q_t)\subseteq\mathcal{D}$ to generate the answer.
Over $B$ queries, the objective is to maximize the fraction of unique retrieved items:
%
%
%
\begin{equation}\label{eq:coverage}
\max_{\{q_1, \cdots, q_B\}}\frac{\left| \bigcup_{t=1}^{B} \mathcal{D}_k(q_t) \right|}{|\mathcal{D}|}
\end{equation}
The budget $B$ models practical constraints such as API cost and the need to limit suspicious activity.
%

\noindent\textbf{Adversary’s Capabilities.}
We assume black box access: the attacker submits adaptive natural language queries and observes only final answers.
They have no access to retrieval scores, identifiers, indices, or intermediate states, and they cannot modify the retriever or poison the corpus.
Following prior work~\cite{cohen2024unleashing,jiang2024rag,wang2025silent}, we assume a coarse topic phrase of the target knowledge base is known, as it is often exposed by product descriptions, onboarding pages, or brief interaction.
We also evaluate a weaker attacker with no topic prior that issues one benign probe query and compresses the response into a short topic phrase using an attacker-side LLM (the probe counts toward the budget; Appendix~\ref{app:no-topic-prior}).

\section{Reducing RAG Crawling to ASCP}
\label{sec:problem-formulation}
We formalize the attacker's goal as \emph{RAG Crawling} and reduce it to ASCP (Sec.~\ref{sec:bg-ac}). In this way, coverage-maximizing adaptive greedy policies can provide a principled benchmark.
We instantiate the ASCP components as follows:

\begin{itemize}[leftmargin=10pt,noitemsep,topsep=0pt]
  \item \textbf{Item universe ($\mathcal{U}$).} We set $\mathcal{U}:=\mathcal{D}$, the attacker-hidden corpus, where each item is an atomic document or passage.
  
  \item \textbf{Action space ($\mathcal{Q}$).} We set $\mathcal{Q}$ to be the set of all valid natural-language queries the attacker may issue to the victim RAG.
  
  \item \textbf{Stochastic outcome (realization $\Phi$ and $O_{\Phi}(q)$).} Let $\Phi$ denote a latent realization of the victim RAG pipeline that {determines} retrieval outcomes (including any nondeterminism and attacker-unknown processing such as rewriting or multi-query retrieval).
  Conditioned on $\Phi$, issuing query $q$ leads the pipeline to use a hidden set of $k$ retrieved items to generate the final answer; we denote this outcome by
  \begin{equation}
      O_{\Phi}(q) := \mathcal{D}^{\Phi}_k(q) \subseteq \mathcal{D}.
  \end{equation}

  \item \textbf{Cost and budget ($c(q),B$).} We assume unit query cost $c(q)=1$, so the attacker can issue at most $B$ queries.
This matches common deployment settings where the service API charges on a per-query basis: each query consumes one billable request regardless of its exact content.

  \item \textbf{Coverage function ($f(S, \Phi)$).} For a set of queries $S\subseteq\mathcal{Q}$ and realization $\Phi$, the coverage function is
\begin{equation}
    f(S,\Phi) := \left|\bigcup_{q\in S} O_{\Phi}(q)\right|\, ,
\end{equation}
which matches adversary's goal (Eq.~\ref{eq:coverage}) up to a constant.

\end{itemize}

%
We next define conditional expected marginal gain (CMG) as follows. CMG is the oracle quantity optimized by adaptive greedy in ASCP.

\begin{definition}[Conditional Expected Marginal Gain in RAG Crawling]
\label{def:cmg}
Let $\psi^{R}_{t-1}$ denote the current \emph{retrieval-level} partial realization (i.e., the executed query--outcome pairs under the latent realization $\Phi$) after $t\!-\!1$ issued queries, and let $S_{t-1}\subseteq \mathcal{Q}$ be the set of queries issued so far.
The \emph{conditional expected marginal gain} (CMG) of a new query $q\in\mathcal{Q}$ given $\psi^{R}_{t-1}$ is
\begin{equation}
\Delta(q \mid \psi^{R}_{t-1})
:= \mathbb{E}_{\Phi \mid \psi^{R}_{t-1}}\!\left[
f(S_{t-1} \cup \{q\}, \Phi) - f(S_{t-1}, \Phi)
\right].
\label{eq:delta-query}
\end{equation}
\end{definition}



We then show that RAG Crawling satisfies the assumptions required by ASCP’s greedy approximation theory.
\begin{theorem}[Adaptive Monotonicity and Submodularity in RAG Crawling]
\label{theorem:monotonicity-submodularity}
The RAG Crawling problem instance satisfies adaptive monotonicity and adaptive submodularity.
\end{theorem}
The proof can be found in Appendix~\ref{appendix:proof}.
Building on Theorem~\ref{theorem:adaptive-greedy} and Theorem~\ref{theorem:monotonicity-submodularity}, we have the following:
\begin{theorem}[Near-optimality guarantee of the adaptive greedy policy for RAG Crawling]
\label{theorem:adaptive-greedy-rag}
The adaptive greedy policy for RAG Crawling, which at each step $t$ selects a query
\begin{equation}
q_t \in \arg\max_{q \in \mathcal{Q}} \Delta(q \mid \psi^{R}_{t-1}),
\label{eq:greedy-policy}
\end{equation}
achieves a $(1-1/e)$-approximation to the optimal expected coverage under the same query budget.
\end{theorem}

\begin{figure}[t]
    \centering
    \includegraphics[width=0.6\linewidth]{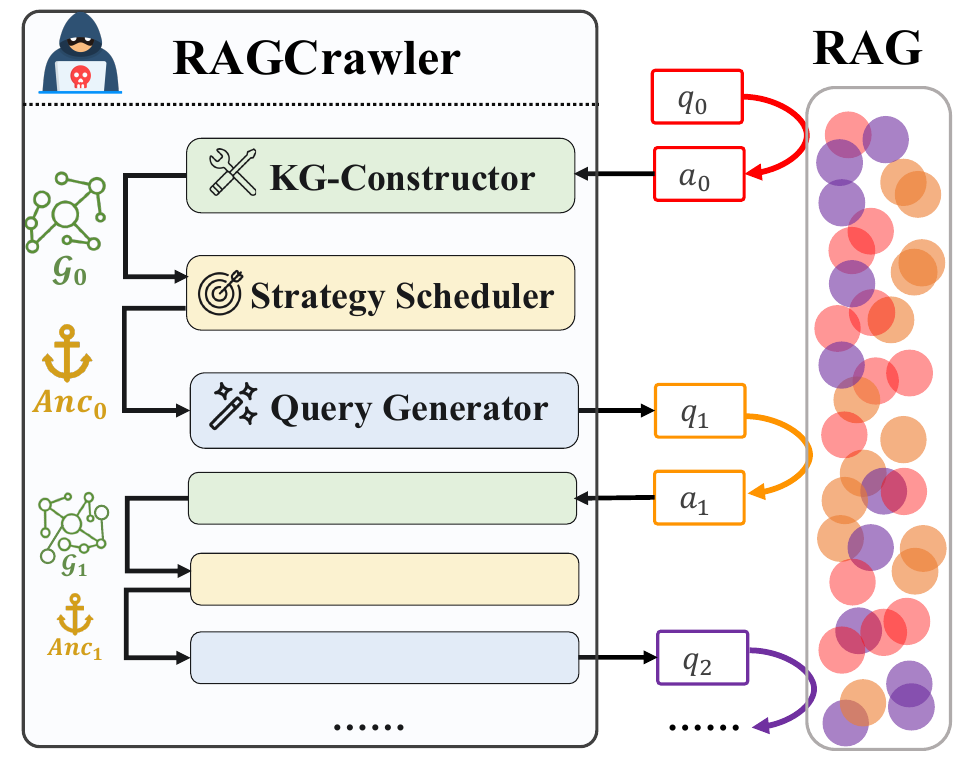}
    \caption{\textbf{The workflow of \toolname.} At each step, (1) The KG-Constructor processes the latest system response to update the knowledge graph. (2) The Strategy Scheduler analyzes this graph to select a strategic anchor. (3) The Query Generator uses this anchor to formulate the next query.}
    \label{fig:overview}
\end{figure}
\section{\toolname}
Theorem~\ref{theorem:adaptive-greedy-rag} provides a theoretical guarantee on optimal coverage of RAG Crawling.  
We now present \toolname, which instantiates the adaptive greedy policy for RAG knowledge-base stealing and enables real-world attacks that achieve 66.8\% coverage on average with only 1,000 queries.



\subsection{Overview}
\label{sec:overview}

In this section, we explain the workflow of \toolname (Fig.~\ref{fig:overview}). \toolname implements a closed-loop crawler that alternates between graph state estimation, anchor scheduling, and natural query realization.


At round $t$, the attacker issues $q_t$ and observes only the answer text $a_t$, while the retrieved documents remain hidden in the black-box setting.
Each answer $a_t$ is processed by \textbf{KG-Constructor} to extract a step-wise knowledge subgraph and update the attacker-side state to $\mathcal{G}_t$. 
Based on $\mathcal{G}_t$, the \textbf{Strategy Scheduler} selects a semantic anchor $Anc_t$ to approximate which direction is likely to yield high \emph{marginal coverage gain} in subsequent queries, where we define $Anc_t \triangleq (e_t^*, r_t^*)$ as an entity with an optional relation target. 
Finally, the \textbf{Query Generator} realizes $Anc_t$ into a fluent and policy-compliant query $q_{t+1}$, sends it to the victim RAG, and the loop repeats. To bootstrap a non-empty graph state, we run a brief topic conditioned seeding phase that generates a few broad queries, and all seeds count toward the budget.

This adaptive loop systematically addresses the three challenges of the ASCP framework. Next, we elaborate on the specific mechanisms employed for each challenge.

\subsection{KG-Constructor}
\label{sec:design-kg}

This module addresses the first challenge in the {ASCP} formulation: \emph{the CMG of the attacker’s action is unobservable}. 
Since the attacker cannot directly observe which documents have been retrieved by the RAG system, the true coverage increment of each query remains hidden. 

\begin{figure}[t]
    \centering
    \includegraphics[width=0.7\linewidth]{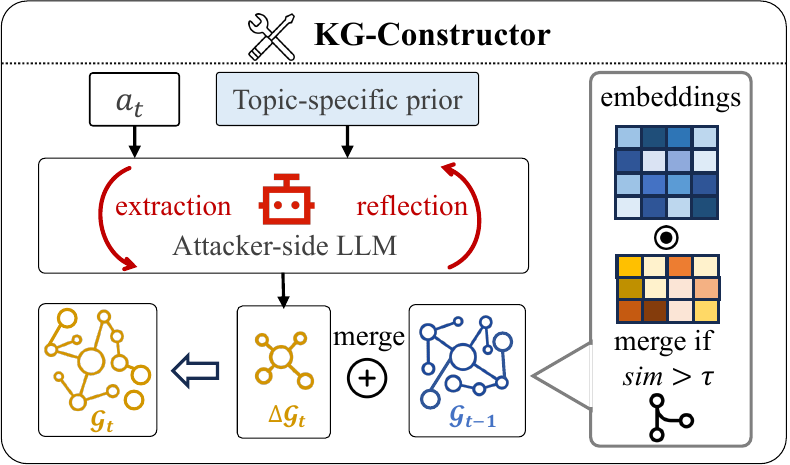}
    \caption{\textbf{KG-Constructor.} Guided by a Topic-Specific Prior, the iterative extraction and reflection process generates a knowledge subgraph ($\Delta\mathcal{G}_t$), which is then refined through incremental graph update to produce the final graph state $\mathcal{G}_t$.}
    \label{fig:kg-constructor}
\end{figure}

KG-Constructor makes coverage gains estimable by converting answer text into an attacker-side knowledge graph that compactly summarizes revealed content. Specifically, we maintain an evolving graph state $\mathcal{G}_t$ that approximates the latent coverage function~$f(S, \Phi)$, making CMG estimable from surface-level answers. The workflow is shown in Fig.~\ref{fig:kg-constructor}.

\noindent{\bf Why existing approaches do not work.}
Existing KG construction pipelines do not fit our setting because we require consistent, lightweight incremental updates from partial and evolving answers, without access to the underlying corpus.
OpenIE-style extraction~\cite{yates2007textrunner} produces free-form relations that drift across rounds, causing redundancy and unstable typing, while schema-constrained methods~\cite{stanovsky2018supervised} reduce variance but typically assume a fixed ontology and full-data access.
LLM-based construction~\cite{papaluca2024zero, bian2025llm} and graph-augmented RAG frameworks (e.g., GraphRAG~\cite{edge2024local}, LightRAG~\cite{guo2024lightrag}) mainly target one-shot document parsing and retrieval quality, and thus do not provide compact, efficient coverage tracking over many rounds of dynamic responses.

\noindent{\bf Our methods.} 
Our KG-Constructor supports efficient incremental maintenance by extracting a per-answer subgraph and merging it into the global graph.
At round $t$, the attacker treats the observed answer $a_t$ as the only evidence and extracts a typed triple set $\{(h,r,t)\}$, where $h$ and $t$ are entity mentions (or attribute values) and $r$ is a semantic relation type.
We build a knowledge delta $\Delta\mathcal{G}_t$ using unique $h/t$ mentions as nodes and triples as directed edges, and merge it into $\mathcal{G}_{t-1}$ to obtain $\mathcal{G}_t$ without rebuilding from scratch.
As shown in Fig.~\ref{fig:kg-constructor}, the workflow has three modules:

\noindent\textbf{1) Topic-Specific Prior.}  
We derive a lightweight topic-specific schema prior to keep extraction focused and consistent across rounds.
Given only a coarse topic signal, we instruct an attacker-side LLM to propose a compact set of entity categories and relation types for the domain (e.g., \textit{disease}, \textit{symptom}, \textit{treatment}, and relations such as \textit{has\_symptom}, \textit{treated\_by}).
During extraction, the LLM prioritizes these schema elements as soft constraints, filtering off-topic content and keeping triples within the topical boundary.
This schema stabilizes relation typing across rounds and keeps prompts compact because the LLM conditions on the abstract schema rather than the full graph, reducing overhead while preserving topical precision.




\noindent\textbf{2) Iterative Extraction and Reflection.} This component uses the attacker-side LLM to process the new answer. To reduce missing edges caused by conservative extraction, we
adopt a multi-round extraction–reflection procedure. As shown in Fig.~\ref{fig:kg-constructor}, each answer is first processed by an ``extraction'' pass, and then reprocessed through a ``reflection'' pass that revisits the same content with awareness of previously omitted facts. This reflection loop enables the LLM to infer implicitly expressed relations and produce additional
triples, forming a more comprehensive knowledge subgraph $\Delta\mathcal{G}_t$.

\noindent\textbf{3) Incremental Graph Update and Semantic Merging.}
We finally update the global graph by merging $\Delta \mathcal{G}_t$ into $\mathcal{G}_{t-1}$ with a lightweight attacker-side operation. 
This integration is implemented as a lightweight merging operation without rebuilding graph from scratch. 
To ensure that the graph structure reflects genuine semantic expansion rather than redundant surface-level variations, we perform a semantic merging stage. 
All entity mentions and relation phrases are encoded into a shared embedding space using a pre-trained encoder. 
Node or edge pairs whose cosine similarity exceeds a threshold~$\tau$ are automatically identified and merged. 

This normalization consolidates synonymous mentions across rounds, keeps $\mathcal{G}_t$ compact and stable for CMG estimation and planning space for {Strategy Scheduler}~(Sec.~\ref{sec:design-sched}).

\subsection{Strategy Scheduler}
\label{sec:design-sched}

This module addresses the second challenge of the {ASCP} formulation: \emph{attacker action space is intractable}. Operating on the knowledge graph from the KG-Constructor, its task is to select the optimal entity-relation pair (e.g., (\diskicon, \textit{HotFix})) to guide the next query. The challenge is the space of potential strategic actions is combinatorially vast, rendering a brute-force search for the optimal pair computationally infeasible. 

\begin{figure}
    \centering
    \includegraphics[width=0.6\linewidth]{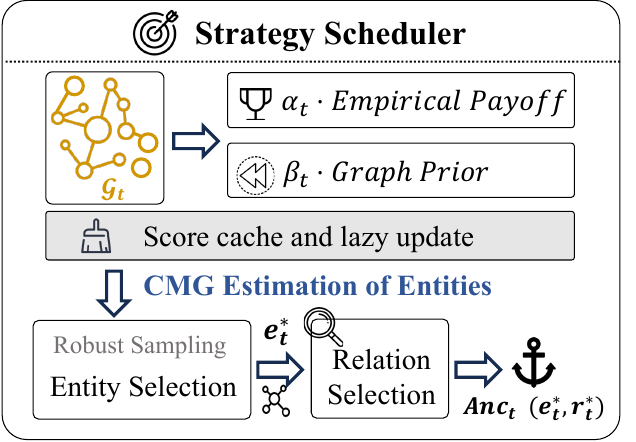}
    \caption{\textbf{Strategy Scheduler.} The scheduler selects an action from the graph $\mathcal{G}_t$ via a two-stage process. First, it samples an entity ($e_t^*$) based on a score that estimates CMG by balancing two key metrics. Second, it selects a relation ($r_t^*$) by identifying the entity's largest local information deficit. A score cache with lazy updates ensures the efficiency of this process.}
    \label{fig:strategy-scheduler}
\end{figure}

Our key insight is \emph{an asymmetry between entities and relations}: entities are the primary drivers of coverage expansion, while relations refine the exploration direction. Exploiting this asymmetry, we reframe the problem into a hierarchical two-stage decision: first select an entity, then pick its unexplored relation to probe. The workflow is shown in Fig.~\ref{fig:strategy-scheduler}.

\noindent{\bf Entities Selection. }We estimate the utility of each entity in $\mathcal{G}_t$ using two complementary metrics: \emph{empirical payoff} and \emph{graph prior}. The \emph{empirical payoff} captures historical information gain and reflects how informative an entity has been in past queries. The \emph{graph prior} exploits the topology of the knowledge graph to identify entities located in structurally under-explored regions.
We combine them into a composite score and sample the anchor entity from a Top-$K$ softmax distribution to preserve diversity under noisy estimates.

\underline{\it Empirical payoff.}
We define the empirical payoff of an entity $e$ via an upper-confidence style score:
\begin{equation}
\mathrm{EmpiricalPayoff}(e) = \bar g_e + c \sqrt{\frac{\log N}{n_e + 1}},
\end{equation}
where $n_e$ is the number of times $e$ has been selected as the anchor, $N$ is the total number of anchor selections so far, and $c>0$ controls the strength of the bonus~\cite{kaufmann2012bayesian,auer2002finite}.

We define $\bar g_e$ as the average normalized graph growth observed when selecting $e$:
\begin{equation}
\bar g_e \;=\; \frac{1}{\max(1,n_e)}\sum_{i \in \mathcal{I}(e)}
\left(
\frac{\Delta \mathcal{E}_i}{\Delta \mathcal{E}_{\max}+\epsilon}
+
\frac{\Delta \mathcal{R}_i}{\Delta \mathcal{R}_{\max}+\epsilon}
\right),
\label{eq:ge}
\end{equation}
where $\mathcal{I}(e)$ is the set of rounds in which $e$ served as the anchor, and $\Delta \mathcal{E}_i$ and $\Delta \mathcal{R}_i$ are the numbers of newly added entities and relation types in round $i$ after semantic merging.
To make gains comparable across rounds, we normalize by window maxima computed over the most recent $t_w$ rounds, denoted by $\mathcal{W}_t$, i.e.,
$\Delta \mathcal{E}_{\max}=\max_{j\in \mathcal{W}_t}\Delta \mathcal{E}_j$ and
$\Delta \mathcal{R}_{\max}=\max_{j\in \mathcal{W}_t}\Delta \mathcal{R}_j$.


\underline{{\it Graph prior.}}
{While the \emph{empirical payoff} provides a self-contained mechanism for balancing exploitation and statistical exploration, it remains blind to the underlying topology of the knowledge graph. Therefore, we introduce a complementary, structure-aware exploration term: the \emph{graph prior}. This prior injects topological intelligence into the strategy, directing it toward graph regions with high discovery potential. It is composed of two distinct terms:}
\begin{equation}
\mathrm{GraphPrior}(e) = \mathrm{DegreeScore}(e) + \mathrm{AdjScore}(e).
\end{equation}

{The first term, $\mathrm{DegreeScore}$, promotes exploration in less-dense regions. It penalizes highly connected entities, operating on the intuition that these ``hub'' nodes are more likely to be information-saturated:}
\begin{equation}
\mathrm{DegreeScore}(e) = 1 - \frac{\deg(e)}{\max_{u\in \mathcal{E}}\deg(u)+\epsilon}.
\end{equation}

{The second term, $\mathrm{AdjScore}$, directly measures an entity's relational deficit. This deficit quantifies how common a specific relation type is among an entity's peers, given that the entity itself lacks that relation. A high $\mathrm{AdjScore}$ thus signals a significant discovery opportunity, prioritizing entities that are most likely to form a new, expected type of connection:}
\begin{equation}
\mathrm{AdjScore}(e) = \max_{r\in \mathcal{R}_t} \mathrm{Deficit}(e,r),
\end{equation}
where the deficit for a specific relation $r$ is defined as:
\begin{equation} 
\resizebox{\linewidth}{!}{$ \mathrm{Deficit}(e,r) = \begin{cases} 
0, & \text{if } r \in \mathrm{EdgeType}(e), \\[6pt] 
\displaystyle 
\frac{1}{|\mathcal{E}_{e.\mathrm{type}}|} \sum_{u\in \mathcal{E}_{e.\mathrm{type}}} \#\mathrm{Edge}(u,r), & \text{if } r \notin \mathrm{EdgeType}(e). \end{cases} $} 
\end{equation}

Here, $\mathcal{R}_t$ denotes the set of relation types currently in the graph $\mathcal{G}_t$, $\mathrm{EdgeType}(e)$ is the set of relation types already connected to $e$, and $\mathcal{E}_{e.\mathrm{type}}$ is the set of all entities sharing the same semantic type as $e$.

\underline{\it Entity scoring and sampling.}
We integrate \emph{empirical payoff} and \emph{graph prior} into a final composite score using time-varying weights:
\begin{equation}
\mathrm{Score}(e) \;=\; \alpha_t \cdot \mathrm{EmpiricalPayoff}(e) \;+\; \beta_t \cdot \mathrm{GraphPrior}(e),
\end{equation}
where $\alpha_t$ gradually increases with time step $t$ to favor high-confidence anchors in later stages, while $\beta_t$ remains positive to maintain structural awareness.

To mitigate the risk of prematurely converging on a seemingly optimal path due to noisy gain estimates, we sample the anchor entity $e_t^*$ from a Top-$K$ softmax distribution over the candidate scores, enhancing strategic diversity and robustness.


\noindent\textbf{Relation Selection.}
Given the sampled anchor $e_t^*$, we choose a relation by probing its largest local deficit:
\begin{equation}
    r^*(e_t^*) \;=\; \arg\max_{r\in \mathcal{R}_t\setminus \mathrm{EdgeType}(e_t^*)}\ \mathrm{Deficit}(e_t^*,r).
\end{equation}

To focus on meaningfully large gaps, we compare this maximum to the global distribution of current deficits,
\begin{equation}
\mathcal{DS}_t \;=\; \big\{\, \mathrm{Deficit}(e,r)\;:\; e\in \mathcal{E}_t,\ r\in \mathcal{R}_t\setminus \mathrm{EdgeType}(e)\,\big\},
\end{equation}
and set $r_t^*=r^*(e_t^*)$ only if it exceeds the $90$th percentile of $\mathcal{DS}_t$; otherwise, we set $r_t^*=\varnothing$ to indicate that no single relation is a sufficiently promising target.


Ultimately, the scheduler outputs the pair $(e_t^*, r_t^*)$, which provides structured guidance to Query Generator (Sec.~\ref{sec:design-qgen}).

\noindent\textbf{Optimization.}
%
We keep scheduling scalable by caching entity scores and lazily recomputing only affected entries after each graph update. 
When $\mathcal{G}_t$ is updated, we invalidate the cache for entities whose neighborhoods changed, and we reuse cached scores for all other entities. 
This reduces per-round overhead and decouples scheduling cost from the full graph size.

\subsection{Query Generator}
\label{sec:design-qgen}

This module addresses the third challenge in the {ASCP} formulation: \emph{Real-world RAG systems impose practical constraints}. We cannot directly send a structured anchor $Anc_t$ to the system, and relying solely on simple templates is not a viable solution, as it can easily be flagged by query provenance mechanisms. Therefore, the core task of this module is to translate the abstract, strategic anchor pair from the {Strategy Scheduler} into a concrete, executable, and natural-sounding query $q_{t+1}$. 

\begin{figure}
    \centering
    \includegraphics[width=0.75\linewidth]{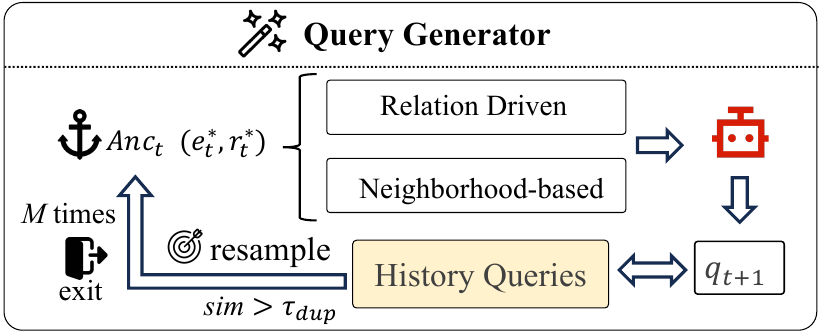}
    \caption{\textbf{Query Generator.} Given an anchor pair $(e_t^*, r_t^*)$, it selects a generation strategy. A candidate query $q_{t+1}$ is produced and checked against historical queries. If redundant, it requests a new anchor from the scheduler and resamples.}
    \label{fig:query-generator}
\end{figure}

This process must address two primary challenges: 1) generating a query that is contextually relevant to the anchor's strategic intent, and 2) ensuring the query remains novel to avoid redundancy under the limited query budget. To achieve this, the generator (Fig.~\ref{fig:query-generator}) involves the following three steps.

\noindent\textbf{1) Adaptive Query Formulation.}
The generator's first step is to intelligently select a formulation strategy based on the specificity of the anchor $(e_t^*, r_t^*)$ provided by the scheduler. This dual-mode approach allows the system to flexibly switch between precision and discovery:
\begin{itemize}[leftmargin=10pt, noitemsep, topsep=0pt]
    \item \textbf{Relation-Driven Probing}. When the {Strategy Scheduler} identifies a plausible relation deficit $r_t^*$ for an entity $e_t^*$ with high confidence, the generator activates this targeted strategy. It instantiates a relation-aware template conditioned on the $(e_t^*, r_t^*)$ pair and realizes it via an LLM into a fluent, natural language query $q_{t+1}$ designed for precise fact-probing.
    \item \textbf{Neighborhood-based Generation.} In the absence of a sufficiently salient relation, the generator shifts to an exploratory mode. It analyzes the local neighborhood of $e_t^*$ within the current knowledge graph $\mathcal{G}_t$. By leveraging the KG's schema and the generative capabilities of LLM, it hypothesizes plausible missing relations and formulates one or more exploratory query variants. 
\end{itemize}

\noindent\textbf{2) History-aware De-duplication}
To maximize the information yield of each query, we introduce a robust quality control gate powered by a history-aware resampling loop. Before being dispatched, each candidate query $q$ is compared against the historical query log $\mathcal{Q}_{\mathrm{hist}}$. If its similarity $\mathrm{sim}(q,\mathcal{Q}_{\mathrm{hist}})\!\ge\!\tau_{dup}$, the generator triggers a resample operation, drawing a new anchor from the {Strategy Scheduler}’s Top-$K$ distribution and re-attempting the formulation, up to $M$ trials. 
This mechanism is critical for avoiding diminishing returns.
If all $M$ trials yield near-duplicate queries, we treat the current exploration as saturated.
{As a practical fallback, an attacker can either restart a seeding step to diversify the query pool~\cite{wang2025silent,jiang2024rag}, or terminate early to avoid wasting the remaining query budget.}

\noindent\textbf{3) Penalties and Feedback Loop.}
We close the loop by feeding generation outcomes back into scheduling. 
When a query is refused, yields an empty answer, or is rejected as a duplicate, we apply a strategy-specific penalty to the corresponding payoff update so that such anchors become less likely to be selected in future rounds. 
This feedback lets the planner adapt away from unproductive regions and toward anchors that reliably produce new graph growth.


\section{Evaluation}
\label{sec:evaluation}
We conduct comprehensive experiments to answer the following research questions:

\noindent\textbf{RQ1:} How effective is \toolname\ compared with existing attack methods?

\noindent\textbf{RQ2:} How does the retriever in the victim RAG affect stealing performance?

\noindent\textbf{RQ3:} How does the attacker-side LLM agent influence stealing performance?

\noindent\textbf{RQ4:} How robust is the attack against RAG variants employing query rewriting or multi-query retrieval?

\noindent\textbf{RQ5:} How do hyperparameters and individual techniques affect the effectiveness of \toolname?

\subsection{Evaluation Setup}
\noindent\textbf{Dataset.}
We evaluate on four corpora spanning diverse domains with varying styles and scales.
TREC-COVID and SciDocs are from BEIR~\cite{thakur2021beir} with 171.3K and 25.6K documents, and NFCorpus~\cite{boteva2016full} with 5.4K documents. We additionally include Healthcare-Magic-100k~\cite{healthcaremagic} (Healthcare) with about 100K patient-provider Q\&A samples.
Together they cover biomedical literature, scientific papers, consumer health, and clinical dialogues.
For efficiency and fair comparison, we randomly sample 1,000 de-duplicated documents from each corpus to form the victim RAG collection~\cite{jiang2024rag,di2024pirates}.
We verify that each subset preserves the full-corpus semantic distribution using Energy Distance~\cite{szekely2013energy} and C2ST~\cite{lopez2016revisiting} (Appendix~\ref{appendix:sample-validation}), and report results with larger victim collections in Appendix~\ref{app:additional-larger}.

\noindent\textbf{Generator and Retriever.} 
We employ two retrievers in our evaluations: BGE~\cite{zhang2023retrieve} and GTE~\cite{li2023towards}.
For generators, we evaluate four large language models: Llama~3.1 Instruct-8B (denoted as Llama-3-8B)~\cite{dubey2024llama}, Command-R-7B~\cite{command-r7b}, Microsoft-Phi-4~\cite{abdin2024phi}, and GPT-4o-mini~\cite{gpt4o-mini}.
These generators cover both open-source and proprietary systems and span different model families and sizes.
Including GPT-4o-mini, which incorporates safety guardrails~\cite{gpt-guardrail}, allows us to evaluate whether stealing remains feasible under built-in guardrails.

\noindent\textbf{Attacker-side LLM.} 
We adopt two attacker-side LLMs to simulate query generation. In the main experiments, we use {Doubao-1.5-lite-32K}~\cite{doubao} due to its high speed, cost efficiency, and independence from the RAG’s generator family (aligns with the black-box assumption). 
We also evaluate a smaller open-source alternative, {Qwen-2.5-7B-Instruct}~\cite{qwen2.5-7b}, to examine transferability between model families and sizes.

\noindent\textbf{RAG Setting.} 
We consider three victim configurations: vanilla RAG, query rewriting, and multi-query retrieval with RRF re-ranking (default: 3 query variants per user request).
We set retrieval depth to $k=10$ by default and analyze the impact of $k$ in Sec.~\ref{sec:ablation}.
The retrieved documents are provided to the generator as contextual input through a structured system prompt. 
Further details are provided in Appendix~\ref{appendix:prompts}.

\noindent\textbf{Baselines.}
We compare \toolname\ against two state-of-the-art black-box RAG knowledge-base stealing attacks: {RAGThief}~\cite{jiang2024rag} and {IKEA}~\cite{wang2025silent}. RAGThief represents continuation-based attack, and IKEA represents keyword-based attack.
To ensure fair comparison, all attacks use the same attacker-side LLM, and we adopt default hyperparameters from the original papers. Unless otherwise specified, each attack is allowed to issue at most 1,000 queries to the victim RAG, ensuring consistent query budgets across methods.

\begin{figure*}[t]
    \centering
    \includegraphics[width=0.8\linewidth]{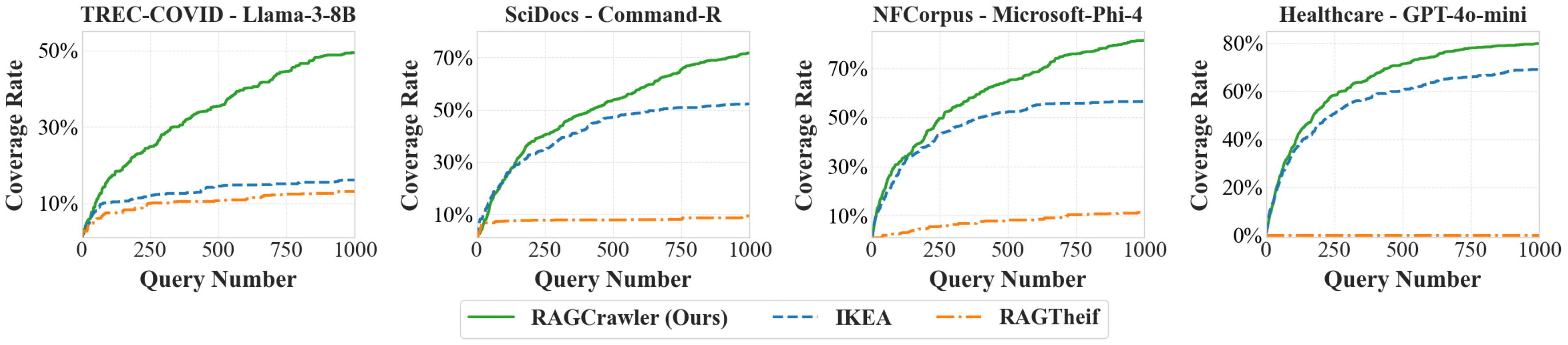}
    \caption{Coverage Rate vs. Query Number (1,000 Budget) across datasets and generators (BGE Retriever). \toolname\ steadily increases coverage and consistently surpasses both RAGThief and IKEA. Full results are in Fig.~\ref{fig:all-coverage}.}
    \label{fig:main-coverage}
\end{figure*}


\noindent\textbf{Metrics.}
We evaluate stealing along four complementary axes: exposure breadth, semantic faithfulness of the stolen content, query cost to reach target coverage, and whether the extracted corpus can support service replication.

\noindent\underline{\it Coverage Rate (CR)} measures the fraction of the corpus exposed through retrieval, matching Eq.~\ref{eq:coverage}; refusals are treated as empty outcomes as in~\cite{jiang2024rag,wang2025silent}.

\noindent\underline{\it Semantic Fidelity (SF)} measures average semantic overlap between each document $d\in\mathcal{D}$ and the stolen content $\mathcal{A}$, where $\mathcal{A}$ is the set of answer snippets produced across all issued queries, using a fixed encoder $E_{\mathrm{eval}}(\cdot)$:
\begin{equation}
    s(d)=\max_{e\in\mathcal{A}} \ \cos\bigl(E_{\mathrm{eval}}(d),\,E_{\mathrm{eval}}(e)\bigr).
\end{equation}
\noindent\underline{\it Target-Coverage Query Complexity} $Q_{\gamma}$ is the minimum number of \emph{issued} queries needed to reach coverage $\gamma$ (the first $t$ such that $\mathrm{CR}(t)\ge \gamma$).
When the target is not reached under a cap $Q_{\max}$, we treat $Q_{\gamma}$ as right-censored and report $Q_{\gamma}>Q_{\max}$.

\noindent\underline{\it Reconstruction Fidelity} assesses whether the stolen corpus supports service replication.
We build a surrogate RAG from the stolen content and evaluate it on official query annotations of TREC-COVID, SciDocs, and NFCorpus, restricted to the sampled documents (1,860, 2,181, and 1,991 queries).
Healthcare is excluded due to the lack of official query annotations.
We report success rate (fraction of non-refusal responses), answer similarity (semantic similarity between surrogate and victim answers), and ROUGE-L.

\subsection{Effectiveness (RQ1)}
\label{sec:effectiveness}

We evaluate the effectiveness of \toolname\ through corpus coverage, semantic fidelity, target-coverage query complexity, and reconstruction fidelity. Across all evaluations, \toolname\ demonstrates the strongest performance.

\begin{table}[t]
\centering

\caption{Coverage rate (CR) and semantic fidelity (SF) of attacks under a 1,000-query budget (victim retriever: BGE) across four datasets and four generators. 
Abbreviations: Gen (Generator), TREC-C (TREC-COVID), Cmd-R (Command-R), MS-Phi (Microsoft-Phi-4), 4o-mini (GPT-4o-mini). 
``N/A'' indicates undefined semantic fidelity due to zero coverage. 
\toolname\ performs best in every setting.}

\label{tab:all-coverage}

\begin{adjustbox}{max width=\linewidth}
\setlength{\tabcolsep}{3.5pt}

\begin{tabular}{llc@{\hspace{5pt}}c@{\hspace{5pt}}cc@{\hspace{5pt}}c@{\hspace{5pt}}c}
\toprule
\multirow{2}{*}{{Dataset}} & \multirow{2}{*}{{Gen}} & \multicolumn{3}{c}{{CR}} & \multicolumn{3}{c}{{SF}} \\
\cmidrule(lr){3-5} \cmidrule(lr){6-8}
 & & RAGThief & IKEA & \textbf{RAGCrawler} & RAGThief & IKEA & \textbf{RAGCrawler} \\
\midrule

\multirow{4}{*}{TREC-C} 
 & Llama      & 0.131 & 0.161 & \textbf{0.494} & 0.447 & 0.495 & \textbf{0.591} \\
 & Cmd-R      & 0.154 & 0.173 & \textbf{0.544} & 0.444 & 0.525 & \textbf{0.614} \\
 & MS-Phi     & 0.121 & 0.197 & \textbf{0.474} & 0.468 & 0.547 & \textbf{0.619} \\
 & 4o-mini    & 0.000 & 0.197 & \textbf{0.465} & N/A     & 0.543 & \textbf{0.572} \\
\midrule

\multirow{4}{*}{SciDocs}    
 & Llama      & 0.053 & 0.513 & \textbf{0.661} & 0.264 & 0.495 & \textbf{0.523} \\
 & Cmd-R      & 0.093 & 0.522 & \textbf{0.717} & 0.295 & 0.514 & \textbf{0.561} \\
 & MS-Phi     & 0.065 & 0.545 & \textbf{0.711} & 0.324 & 0.534 & \textbf{0.563} \\
 & 4o-mini    & 0.000 & 0.421 & \textbf{0.516} & N/A     & 0.475 & \textbf{0.492} \\
\midrule

\multirow{4}{*}{NFCorpus}   
 & Llama      & 0.061 & 0.503 & \textbf{0.797} & 0.451 & 0.644 & \textbf{0.698} \\
 & Cmd-R      & 0.169 & 0.517 & \textbf{0.844} & 0.467 & 0.656 & \textbf{0.705} \\
 & MS-Phi     & 0.113 & 0.566 & \textbf{0.813} & 0.487 & 0.663 & \textbf{0.717} \\
 & 4o-mini    & 0.000 & 0.493 & \textbf{0.631} & N/A     & 0.631 & \textbf{0.653} \\
\midrule

\multirow{4}{*}{Healthcare} 
 & Llama      & 0.361 & 0.687 & \textbf{0.807} & 0.536 & 0.588 & \textbf{0.618} \\
 & Cmd-R      & 0.170 & 0.592 & \textbf{0.766} & 0.383 & 0.578 & \textbf{0.582} \\
 & MS-Phi     & 0.052 & 0.588 & \textbf{0.654} & 0.434 & 0.558 & \textbf{0.599} \\
 & 4o-mini    & 0.000 & 0.693 & \textbf{0.799} & N/A     & 0.490 & \textbf{0.577} \\
\midrule

\multicolumn{2}{c}{{Average}} 
 & 0.096 & 0.461 & \textbf{0.668} & 0.417 & 0.559 & \textbf{0.605} \\
\bottomrule
\end{tabular}
\end{adjustbox}
\end{table}

\noindent\textbf{Coverage Rate.}
Under a fixed 1,000-query budget with the BGE retriever, \toolname\ attains the highest final coverage in every configuration in Table~\ref{tab:all-coverage}.
Averaged over all 16 configurations, \toolname\ exposes 66.8\% of the corpus, compared with 46.1\% for IKEA and 9.6\% for RAGThief.
The margin is largest on harder corpora (TREC-COVID), where \toolname\ reaches 49.4\% average coverage across generators, while IKEA remains at 18.2\%. 
The coverage curves in Fig.~\ref{fig:main-coverage} further indicate that \toolname\ continues to accrue new documents throughout the process, whereas baselines typically flatten earlier, consistent with diminishing returns from locally reactive querying.

\toolname\ also remains effective under generators with built-in guardrails.
When the victim generator is GPT-4o-mini, RAGThief yields zero coverage across all four corpora, while \toolname\ still reaches 46.5\%-79.9\% coverage, suggesting that benign, policy-compliant probing can sustain stealing even when overt jailbreak patterns are blocked.

\noindent\textbf{Semantic Fidelity.}
Higher coverage does not come at the expense of semantic quality. 
Across all dataset--generator pairs, \toolname\ achieves the highest average SF (0.605), exceeding IKEA (0.559) and RAGThief (0.417) in Table~\ref{tab:all-coverage}.


\begin{table}[t]
\centering

\caption{Target-coverage query complexity to reach 70\% coverage ($Q_{0.7}$) under a cap $Q_{\max}{=}5{,}000$ (victim: Llama-3-8B generator, BGE retriever).  
Entries marked ``$>Q_{\max}$'' are right-censored, with the achieved coverage shown in parentheses. The full sweep of $Q_{\gamma}$ is reported in Tab.~\ref{tab:qtau}.
\toolname\ reaches high coverage with far fewer queries.}

\label{tab:q70}

\setlength{\tabcolsep}{8pt}

\begin{adjustbox}{max width = 0.9\columnwidth}
\begin{tabular}{lccc}
\toprule
{Dataset} & RAGThief & IKEA & \textbf{RAGCrawler} \\
\midrule
TREC-COVID & $>5{,}000$ (29.8\%) & $>5{,}000$ (19.6\%) & \textbf{4,040} \\
SciDocs    & $>5{,}000$ (7.2\%)  & $>5{,}000$ (56.0\%) & \textbf{1,163} \\
NFCorpus   & $>5{,}000$ (16.5\%) & $>5{,}000$ (64.0\%) & \textbf{595}   \\
Healthcare & $>5{,}000$ (53.0\%) & 1,231             & \textbf{568}   \\
\bottomrule
\end{tabular}
\end{adjustbox}
\end{table}

\noindent\textbf{Target-Coverage Query Complexity.}
We next evaluate query efficiency for reaching high coverage targets.
Table~\ref{tab:q70} reports $Q_{0.7}$ under a cap $Q_{\max}=5{,}000$.
\toolname\ is the only method that reaches 70\% coverage on all datasets within the cap, requiring 4{,}040 queries on TREC-COVID and at most 1{,}163 queries on the other corpora (595 on NFCorpus and 568 on Healthcare).
IKEA reaches 70\% only on Healthcare (1{,}231 queries) and stays below 70\% elsewhere even at $Q_{\max}$, while RAGThief never reaches 70\%.
Relative to the strongest baseline per dataset, \toolname\ reduces $Q_{0.7}$ by 2.17$\times$ on Healthcare and by at least 1.24$\times$, 4.30$\times$, and 8.40$\times$ on TREC-COVID, SciDocs, and NFCorpus (lower bounds due to censoring). This translates to average $\ge 4.03\times$ reduction in monetary cost under linear per-request pricing (e.g., \$20 per 1{,}000 search requests on Vertex AI Search~\cite{vertexaisearch_pricing}).

The full sweep in Tab.~\ref{tab:qtau} highlights the gap. While IKEA can be competitive at low targets on easier corpora (e.g., Healthcare $Q_{0.5}=212$ vs. 209), it saturates as $\gamma$ increases and fails to reach 90\% on any dataset within $Q_{\max}$.
In contrast, \toolname\ reaches 90\% on SciDocs, NFCorpus, and Healthcare with 3{,}330, 1{,}753, and 2{,}434 queries, respectively, and attains 72.5\% final coverage on TREC-COVID, more than doubling the best baseline's (29.8\% for RAGThief).

\begin{table}[t]

\caption{Reconstruction fidelity of surrogate RAG systems built from \emph{stolen} knowledge-base content (victim: Llama-3-8B generator, BGE retriever). 
We report success rate (non-refusal), answer embedding similarity, and ROUGE-L against victim outputs. 
Best results are in \textbf{bold}. 
Surrogates built from \toolname's stolen corpus match the victim best.}

\label{tab:reconstruction}
\centering
\begin{adjustbox}{max width = 0.85\columnwidth}
\begin{tabular}{llc@{\hspace{5pt}}c@{\hspace{5pt}}c}
\toprule
         {Dataset}                   & {Method}     & {Success Rate} & {Answer Sim.} & {ROUGE-L}        \\ \midrule
\multirow{3}{*}{TREC-COVID} & RAGThief   & 0.1129          & 0.4920          & 0.1547          \\
                            & IKEA       & 0.2247          & 0.4779          & 0.1730          \\
                            & \textbf{RAGCrawler} & \textbf{0.3839} & \textbf{0.6098} & \textbf{0.2408} \\ \midrule
\multirow{3}{*}{SciDocs}    & RAGThief   & 0.0486          & 0.4275          & 0.1131          \\
                            & IKEA       & 0.0179          & 0.4829          & 0.1277          \\
                            & \textbf{RAGCrawler} & \textbf{0.3810} & \textbf{0.5900} & \textbf{0.2285} \\ \midrule
\multirow{3}{*}{NFCorpus}   & RAGThief   & 0.0447          & 0.5156          & 0.1063          \\
                            & IKEA       & 0.4215          & 0.6064          & 0.2013          \\
                            & \textbf{RAGCrawler} & \textbf{0.5259} & \textbf{0.6992} & \textbf{0.2334} \\ \bottomrule
\end{tabular}
\end{adjustbox}
\end{table}

\noindent\textbf{Reconstruction Fidelity.}
To test whether the stolen corpus enables service replication, we build surrogate RAG systems from each method's stolen outputs and evaluate them on held-out query sets.
Table~\ref{tab:reconstruction} shows that surrogates built from \toolname\ achieve substantially higher success rates (38.1\%-52.6\%) and consistently higher answer similarity and ROUGE-L, with peaks of 0.6992 similarity and 0.2408 ROUGE-L.
Notably, on SciDocs the IKEA-based surrogate has a very low success rate (0.0179) despite IKEA reaching 51.3\% coverage under the 1,000-query budget in Table~\ref{tab:all-coverage}, indicating that partial theft concentrated in a limited region does not necessarily translate into functional replication.
Overall, the reconstruction results align with the coverage and fidelity trends: broader exploration coupled with higher semantic alignment yields stolen corpora that are materially more useful for building a substitute service.

\noindent\textbf{Takeaway.} Across datasets and generators, \toolname\ delivers the strongest stealing breadth, semantic fidelity, and query efficiency, and the resulting stolen content transfers into markedly higher-fidelity surrogate reconstruction.

\subsection{Retriever Sensitivity (RQ2)}
\label{sec:retriever}

To evaluate the robustness of \toolname\ against different retrieval architectures, we evaluated the attack methods on victim RAG systems with GTE~\cite{li2023towards} as retriever. %

Swapping the retriever shifts the attack surface for all methods, but \toolname\ remains the most effective extractor in terms of coverage.
With GTE, \toolname\ achieves 76.6\% average coverage across datasets, compared with 53.9\% for IKEA and 28.2\% for RAGThief (Table~\ref{tab:gte-coverage}).
The effect of the retriever is not uniform across corpora: relative to BGE, \toolname\ gains 27.1 points on TREC-COVID (49.4\% to 76.5\%) and 17.2 points on SciDocs (66.1\% to 83.3\%), but drops by 5.9 and 7.9 points on NFCorpus and Healthcare, respectively.
This variation suggests that retriever choice can materially change which parts of a corpus are easy to surface, yet \toolname\ consistently maintains high coverage (at least 72.8\%) across all four corpora under GTE.


\begin{table}[t]
\centering
\caption{Coverage rate (CR) and semantic fidelity (SF) of attacks when the victim uses \textbf{GTE} as retriever (generator: Llama-3-8B; 1,000-query budget). 
Best results are in \textbf{bold}. \toolname\ remains dominant.}
\label{tab:gte-coverage}

\begin{adjustbox}{max width=0.95\columnwidth}
\setlength{\tabcolsep}{3.5pt}
\begin{tabular}{lc@{\hspace{5pt}}c@{\hspace{5pt}}cc@{\hspace{5pt}}c@{\hspace{5pt}}c}
\toprule
\multirow{2}{*}{{Dataset}} & \multicolumn{3}{c}{{CR}} & \multicolumn{3}{c}{{SF}} \\
\cmidrule(lr){2-4} \cmidrule(lr){5-7}
 & RAGThief & IKEA & \textbf{RAGCrawler} & RAGThief & IKEA & \textbf{RAGCrawler} \\
\midrule
TREC-COVID  & 0.191 & 0.391 & \textbf{0.765} & 0.453 & 0.565 & \textbf{0.610} \\
SciDocs     & 0.162 & 0.472 & \textbf{0.833} & 0.345 & 0.495 & \textbf{0.539} \\
NFCorpus    & 0.386 & 0.622 & \textbf{0.738} & 0.589 & 0.648 & \textbf{0.671} \\
Healthcare  & 0.388 & 0.671 & \textbf{0.728} & 0.548 & \textbf{0.576} & {0.567} \\
\midrule
{Average} & 0.282 & 0.539 & \textbf{0.766} & 0.484 & 0.571 & \textbf{0.597} \\
\bottomrule
\end{tabular}
\end{adjustbox}
\end{table}

Semantic fidelity remains broadly stable. 
Averaged over datasets, \toolname\ attains SF 0.597, exceeding IKEA (0.571) and RAGThief (0.484). The only exception is Healthcare, where IKEA achieves marginally higher SF (0.576 vs.\ 0.567), indicating that retriever changes can also alter per-document alignment quality in specific domains.

\noindent\textbf{Takeaway.}
\toolname\ generalizes across retrievers: changing the victim retriever can affect absolute difficulty and the coverage--fidelity balance, but \toolname\ remains the strongest overall threat, especially in coverage.



\begin{table}[b]
\centering
\caption{Coverage rate (CR) and semantic fidelity (SF) of attacks when using Qwen-2.5-7B-Instruct as the attacker-side agent (victim: Llama-3-8B generator, BGE retriever; 1,000-query budget). 
Best results are in \textbf{bold}. 
\toolname\ remains dominant.}
\label{tab:qwen-coverage}

\begin{adjustbox}{max width=0.95\columnwidth}
\setlength{\tabcolsep}{3.5pt}
\begin{tabular}{lc@{\hspace{5pt}}c@{\hspace{5pt}}cc@{\hspace{5pt}}c@{\hspace{5pt}}c}
\toprule
\multirow{2}{*}{{Dataset}} & \multicolumn{3}{c}{{CR}} & \multicolumn{3}{c}{{SF}} \\
\cmidrule(lr){2-4} \cmidrule(lr){5-7}
 & RAGThief & IKEA & \textbf{RAGCrawler} & RAGThief & IKEA & \textbf{RAGCrawler} \\
\midrule
TREC-COVID  & 0.271 & 0.183 & \textbf{0.542} & 0.469 & 0.499 & \textbf{0.570} \\
SciDocs     & 0.099 & 0.489 & \textbf{0.675} & 0.298 & 0.487 & \textbf{0.507} \\
NFCorpus    & 0.314 & 0.559 & \textbf{0.834} & 0.559 & 0.646 & \textbf{0.678} \\
Healthcare  & 0.414 & 0.687 & \textbf{0.799} & 0.549 & 0.576 & \textbf{0.584} \\
\midrule
{Average} & 0.275 & 0.480 & \textbf{0.713} & 0.469 & 0.552 & \textbf{0.585} \\
\bottomrule
\end{tabular}
\end{adjustbox}
\end{table}

\subsection{Agent Sensitivity (RQ3)}
\label{sec:agent}
To investigate the influence of the attacker's LLM agent on attack performance, we evaluated the effectiveness with a smaller open-source model, Qwen-2.5-7B-Instruct.

\toolname\ remains effective with a smaller attacker model.
With Qwen-2.5-7B-Instruct (victim: Llama-3-8B, BGE), \toolname\ achieves 71.3\% average coverage with 0.585 semantic fidelity (Table~\ref{tab:qwen-coverage}).
Compared to the Doubao agent under the same victim setting, the per-dataset coverage changes are small (within a few percentage points), suggesting that the global scheduler and graph state dominate the attack trajectory, while the LLM agent primarily executes localized, well-specified generation tasks.

In comparison, the baseline methods remain significantly less effective. With the Qwen-2.5-7B-Instruct agent, IKEA achieves an average coverage of 48.0\%, while RAGThief reaches 27.5\%. Although RAGThief sees a slight performance improvement (likely because the constrained divergent ability of smaller model make it less prone to the off-topic drift), it still lags far behind \toolname.

\noindent\textbf{Takeaway.} Attack effectiveness for \toolname\ is largely agent-agnostic, reinforcing that its advantage comes from planning and redundancy control rather than relying on a high-capability attacker LLM.

\begin{table}[t]
\centering

\caption{Coverage rate (CR) and semantic fidelity (SF) under practical RAG variants (victim: Llama-3-8B generator, BGE retriever, 1,000-query budget).
Best results are in \textbf{bold}. \toolname\ remains the strongest attack.}

\label{tab:defense-robustness-stacked}
\scriptsize
\setlength{\tabcolsep}{3.2pt}
\begin{adjustbox}{max width=\columnwidth}
\begin{tabular}{lccc ccc}
\toprule
\multicolumn{7}{c}{\textbf{Query Rewriting}} \\
\midrule
\multirow{2}{*}{{Dataset}} & \multicolumn{3}{c}{{CR}} & \multicolumn{3}{c}{{SF}} \\
\cmidrule(lr){2-4} \cmidrule(lr){5-7}
& RAGThief & IKEA & \textbf{RAGCrawler} & RAGThief & IKEA & \textbf{RAGCrawler} \\
\midrule
TREC-COVID & 0.381 & 0.241 & \textbf{0.601} & 0.519 & 0.537 & \textbf{0.591} \\
SciDocs    & 0.561 & 0.542 & \textbf{0.743} & 0.442 & 0.479 & \textbf{0.507} \\
NFCorpus   & 0.664 & 0.489 & \textbf{0.854} & 0.633 & 0.618 & \textbf{0.687} \\
Healthcare & 0.709 & 0.595 & \textbf{0.767} & 0.556 & 0.564 & \textbf{0.577} \\
\midrule
{Average}     & 0.579 & 0.467 & \textbf{0.741} & 0.538 & 0.550 & \textbf{0.591} \\
\midrule
\multicolumn{7}{c}{\textbf{Multi-query Retrieval}} \\
\midrule
\multirow{2}{*}{{Dataset}} & \multicolumn{3}{c}{{CR}} & \multicolumn{3}{c}{{SF}} \\
\cmidrule(lr){2-4} \cmidrule(lr){5-7}
& RAGThief & IKEA & \textbf{RAGCrawler} & RAGThief & IKEA & \textbf{RAGCrawler} \\
\midrule
TREC-COVID & 0.326 & 0.189 & \textbf{0.474} & 0.525 & 0.523 & \textbf{0.581} \\
SciDocs    & 0.352 & 0.508 & \textbf{0.697} & 0.364 & 0.494 & \textbf{0.514} \\
NFCorpus   & 0.392 & 0.540 & \textbf{0.849} & 0.588 & 0.631 & \textbf{0.692} \\
Healthcare & 0.562 & 0.605 & \textbf{0.761} & 0.569 & 0.580 & \textbf{0.587} \\
\midrule
{Average}    & 0.408 & 0.461 & \textbf{0.695} & 0.512 & 0.557 & \textbf{0.594} \\
\bottomrule
\end{tabular}
\end{adjustbox}
\end{table}

\subsection{Defense Robustness (RQ4)}
\label{sec:defense}

To further explore potential defenses, we extend our evaluation beyond the test against GPT-4o-mini's built-in guardrails (Sec.~\ref{sec:effectiveness}) to two practical mechanisms widely adopted by RAG: query rewriting and multi-query retrieval. 

\noindent\textbf{Query Rewriting.}
In this configuration, the victim RAG system employs an LLM to rewrite incoming queries, aiming to clarify user intent and neutralize potential adversarial patterns before retrieval and generation. \toolname\ exhibits exceptional resilience against this defense, maintaining superior performance metrics. As detailed in Table~\ref{tab:defense-robustness-stacked}, \toolname\ achieves an average coverage rate of 74.1\% and semantic fidelity of 0.591. Although intended as a safeguard, the query rewriting process is paradoxically exploited by our method to enhance theft. By explicitly refining the query’s semantic intent, the rewriter enables the retriever to surface a more relevant and diverse set of documents than the original input would yield. \toolname's adaptive planner capitalizes on this context to accelerate corpus exploration; for instance, on the NFCorpus dataset, coverage increases from 79.7\% to 85.4\% when this defense is active.

In contrast, baseline methods fail to effectively exploit this dynamic. While RAGThief sees its average coverage improve to 57.9\% and IKEA reaches 46.7\%, both remain significantly behind \toolname. RAGThief benefits incidentally, as the rewriting step strips away its obvious adversarial suffixes; however, lacking a strategic framework to systematically capitalize on the enhanced retrieval results, it cannot match \toolname's efficiency, underscoring the fundamental limitations of heuristic-based approaches.

\noindent\textbf{Multi-query Retrieval.}
We next evaluate against multi-query retrieval, which expands each query into several variants whose retrieved results are re-ranked and fused, a process that can influence attack surface. Once again, \toolname\ excels, achieving the highest average coverage of 69.5\% and a semantic fidelity of 0.593 (Table~\ref{tab:defense-robustness-stacked}). This mechanism provides the attacker with a more diverse set of retrieved documents from different semantic clusters. The global graph in \toolname\ is uniquely positioned to exploit this; it integrates this diverse information to build a more comprehensive map of the corpus, thereby generating more effective and far-reaching follow-up queries.

The baseline methods, however, are not equipped to fully leverage this enriched context.  RAGThief and IKEA attain coverage rates of 40.8\% and 46.1\%, respectively. Their localized strategies are unable to fully synthesize the information from the multiple retrieval results.

\noindent\textbf{Takeaway.}
\toolname\ remains highly effective against practical RAG defenses. 
The results demonstrate that defenses designed to sanitize inputs or strengthen retrieval can be subverted by a strategic attacker and may even amplify the stealing threat, calling for more advanced safeguards.

\subsection{Ablation Study (RQ5)}
\label{sec:ablation}

We analyze how attacker assumptions, efficiency techniques, and victim-side hyperparameters influence \toolname's effectiveness. More ablations on modules and hyperparameter choices are provided in Appendix~\ref{appendix:hyper-parameter}.

\begin{table}[t]
\caption{Coverage rate (CR) with a provided topic prior vs.\ a no-prior attacker (victim: Llama-3-8B generator, BGE retriever). Best results are in \textbf{bold}. \toolname\ remains best and shows small sensitivity to topic-prior removal.}

\label{tab:no-topic-prior}
\centering
\small
\setlength{\tabcolsep}{4pt}
\begin{adjustbox}{max width = \columnwidth}
\begin{tabular}{lcccccccc}
\toprule
& \multicolumn{2}{c}{TREC-COVID} & \multicolumn{2}{c}{SciDocs} & \multicolumn{2}{c}{NFCorpus} & \multicolumn{2}{c}{Healthcare} \\
\cmidrule(lr){2-3}\cmidrule(lr){4-5}\cmidrule(lr){6-7}\cmidrule(lr){8-9}
Method & Prior & No-prior & Prior & No-prior & Prior & No-prior & Prior & No-prior \\
\midrule
RAGThief & 0.131 & 0.128 & 0.053 & 0.040 & 0.061 & 0.123 & 0.361 & 0.290 \\
IKEA     & 0.161 & 0.391 & 0.513 & 0.108 & 0.503 & 0.536 & 0.687 & 0.705 \\
\textbf{RAGCrawler} & \textbf{0.494} & \textbf{0.426} & \textbf{0.661} & \textbf{0.613} & \textbf{0.797} & \textbf{0.825} & \textbf{0.807} & \textbf{0.755} \\
\bottomrule
\end{tabular}
\end{adjustbox}
\end{table}

\noindent\textbf{Robustness to topic prior.}
Our main threat model assumes a coarse topic phrase~\cite{cohen2024unleashing,jiang2024rag,wang2025silent}, which is typically exposed by production RAG services. We further evaluate a weaker attacker that starts with no topic prior and infers a coarse 3--8 word topic phrase using a single benign probe query (counted toward the budget).
Table~\ref{tab:no-topic-prior} shows that \toolname remains the best-performing attack in this setting, with only modest CR changes relative to the topic-prior setting. Appendix~\ref{app:no-topic-prior} details the one-shot probing procedure and analyzes the quality and variability of the inferred topic phrases.

\begin{figure}[t]
    \centering
    \includegraphics[width=0.8\linewidth]{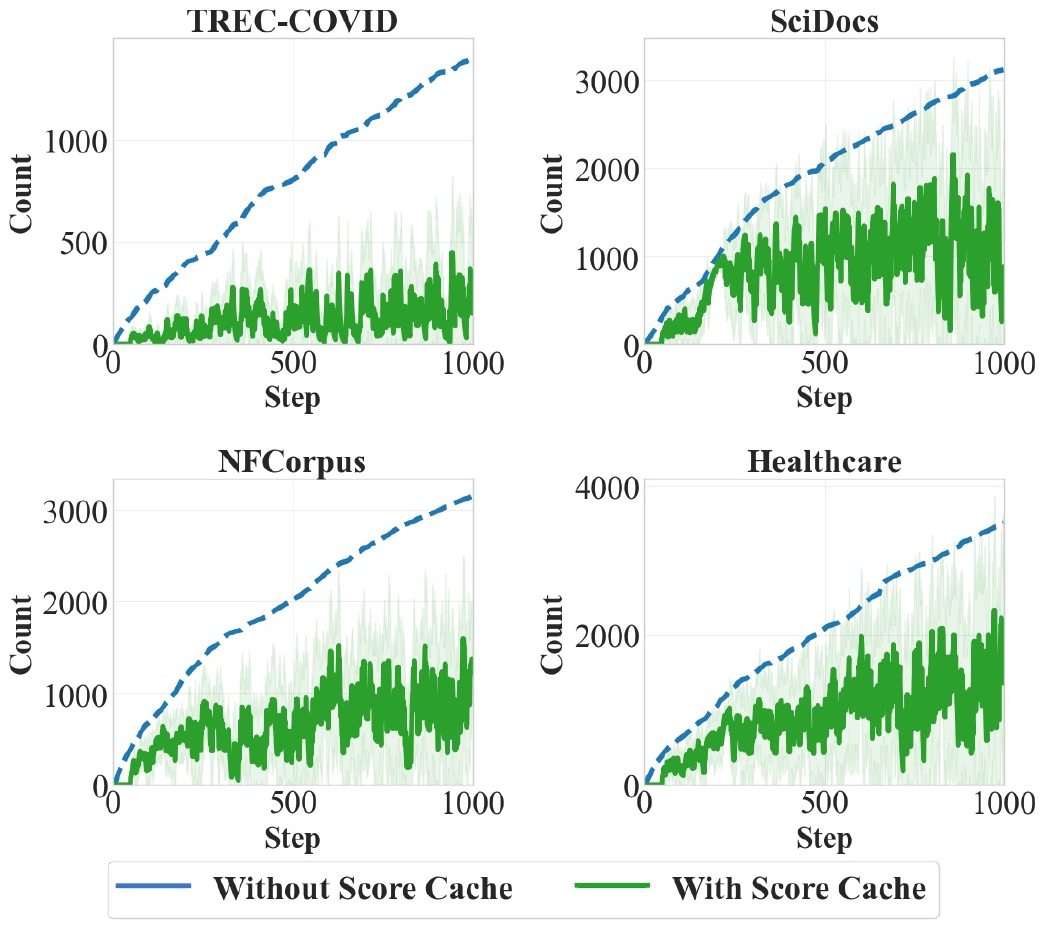}
    \caption{Score Computations with vs. without Caching. Caching reduces the number of similarity score updates by $\sim$2.67$\times$ on average, greatly improving efficiency.}
    \label{fig:score-cache}
\end{figure}

\noindent\textbf{Score Cache.}
We quantify the computational benefit of caching similarity-score updates inside the strategy scheduler.
Figure~\ref{fig:score-cache} shows that caching reduces the number of score updates from over 1.67 million to about 0.63 million across datasets, corresponding to a 62\% workload reduction (2.67$\times$ fewer updates).
Because later rounds increasingly revisit overlapping neighborhoods in the evolving graph, the cache benefit grows with interaction length, improving practicality for larger budgets or larger corpora.

\begin{figure}[b]
    \centering
    \includegraphics[width=0.7\linewidth]{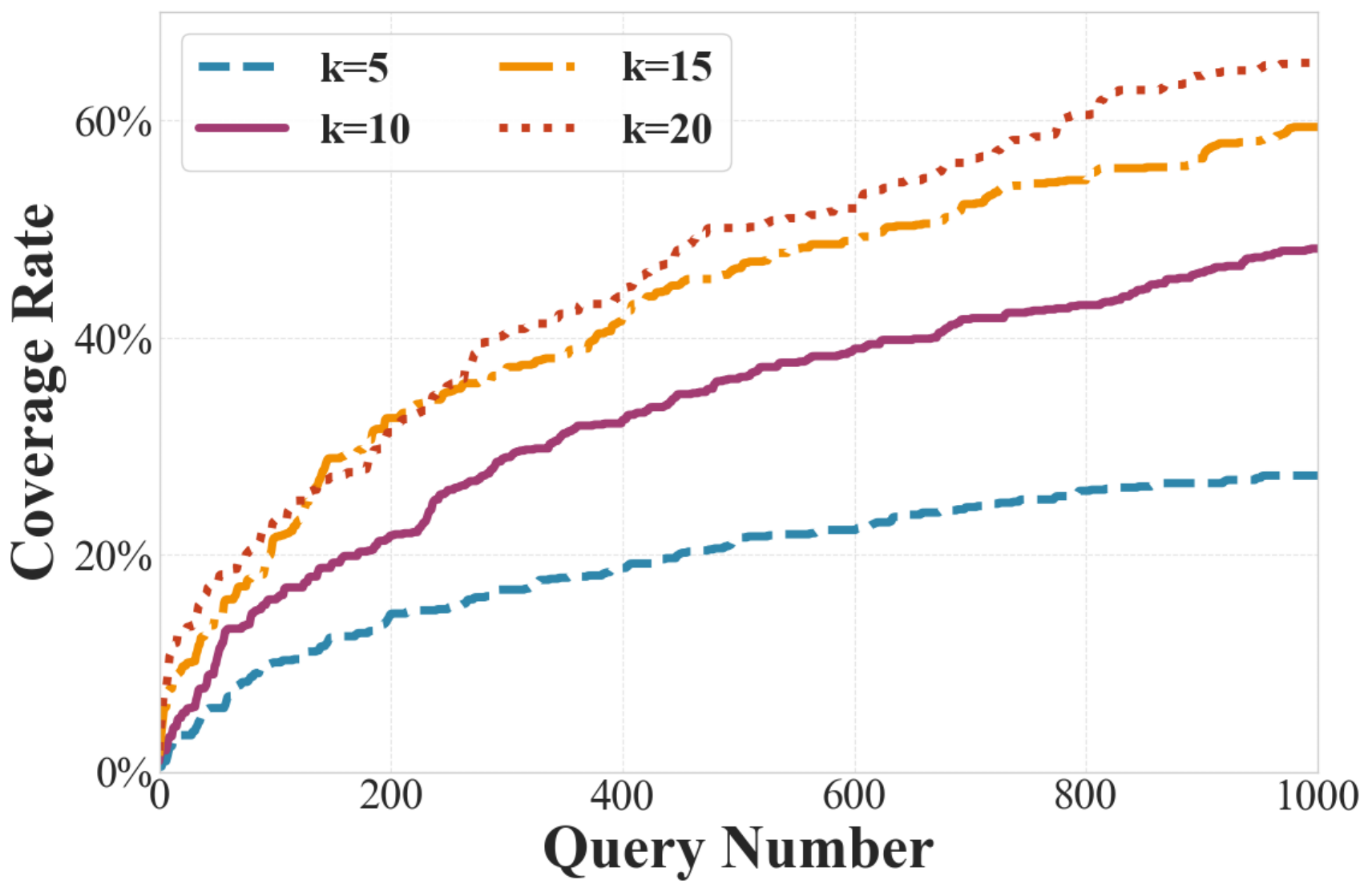}
    \caption{Coverage Rate vs. Query Number for different victim depth $k$ on TREC-COVID, Llama-3-8B, BGE.}
    \label{fig:retrieve_num}
    
\end{figure}

\noindent\textbf{Victim Retrieval Depth.}
We vary the victim retrieval depth $k$ to study how much per-query context expansion affects attack performance.
On TREC-COVID, increasing $k$ substantially accelerates early coverage growth: within 1,000 queries, coverage rises from roughly 28\% at $k=5$ to above 60\% at $k=20$ (Fig.~\ref{fig:retrieve_num}).
However, the curve exhibits diminishing returns as $k$ increases, consistent with additional retrieved documents being increasingly redundant with already exposed content.
This result highlights a security implication of aggressive retrieval: larger $k$ increases the amount of unique corpus material that can be exposed per user query.

\section{Discussion and Related Work}
\label{sec:discussion}

\noindent\textbf{Cost Analysis.}
We separate (i) attacker-side {orchestration} cost (tokens for running \toolname) from (ii) victim-side {service} cost (per-query fees).
Tab.~\ref{tab:cost} shows \toolname incurs only \$0.33--\$0.53 per dataset under Doubao-1.5-lite-32K pricing, and becomes \textbf{near-zero} with open-source models (Sec.~\ref{sec:agent}) aside from GPU hosting.
Victim-side fees scale linearly with issued queries (e.g., \(Q_{0.7}\) in Tab.~\ref{tab:q70}). The low attacker overhead creates strong economic asymmetry against RAG development cost~\cite{adasci2024ragcost}, enabling low-cost service piracy.

\noindent{\textbf{Scalability.}}
Although production RAG corpora may contain millions of documents, \toolname\ scales because each turn extracts a compact typed delta from the {single} observed answer to update the attacker-side KG, keeping per-turn LLM cost independent of \(|\mathcal{D}|\).
The main bottleneck is accumulated-state maintenance (semantic merging and score updates), which we mitigate with caching and lazy recomputation (62\% fewer score updates; Fig.~\ref{fig:score-cache}).
For larger budgets, ANN-backed deduplication~\cite{malkov2018efficient,yang2025effective} and out-of-core graph and score storage~\cite{ai2017squeezing,zhang2018wonderland,xu2020hybrid} offer orthogonal engineering extensions.

\noindent\textbf{Defenses.}
Our attack highlights the inherent limitations of defenses that rely on analyzing queries in isolation (e.g., guardrails or query rewriting). {Moreover, when we evaluate an LLM-based intent detector, it flags only \textbf{1.07\%} of the attack queries as suspicious.} The core challenge is that the malicious pattern is an emergent property of the entire interaction sequence, not an attribute of any single query. This exposes a critical blind spot in static security models and argues for a pivot towards dynamic, behavior-aware defenses.
Based on this analysis, we believe query provenance analysis~\cite{li2025query,han2020unicorn,li2023nodlink,milajerdi2019holmes} holds significant promise. 
Verifying its practical effectiveness would be a valuable next step for the research community.

\noindent\textbf{Security Risks in RAG.} This paper concentrates on knowledge-base stealing attacks against RAG systems~\cite{qifollow, jiang2024rag, wang2025silent, di2024pirates, cohen2024unleashing}.
RAG systems are also susceptible to other security risks~\cite{lasso2024ragsecurity,ni2025towards2}. Membership Inference Attacks (MIA) enable an adversary to determine if a specific document is present in the corpus, thereby exposing the existence of sensitive records~\cite{liu2025mask, naseh2025riddle, anderson2025my, feng2025ragleak}. Furthermore, in Corpus Poisoning attacks, an adversary injects malicious data into the RAG knowledge base~\cite{zou2025poisonedrag, chaudhari2024phantom, cho2024typos, shafran2024machine,zhong2023poisoning,liu2025poisoned,ben2024gasliteing,liang2025graphrag}. Retrieval of this corrupted data can manipulate the system's output, allowing the attacker to propagate misinformation or harmful content.

\section{Conclusion}

We investigated knowledge-base stealing in RAG systems, a growing threat to the intellectual property of the retrieval corpus.
We formalized the attacker’s goal as \emph{RAG Crawling}, an instance of ASCP that targets {global} corpus coverage under a fixed query budget, and we developed \toolname, a knowledge graph-guided framework that operationalizes this objective in practical black-box settings. 
Extensive experiments show that \toolname achieves substantially higher coverage than prior attacks (up to 84.4\% within a budget of 1,000 queries) and remains effective under query rewriting and multi-query retrieval, highlighting urgent gaps in protecting the knowledge assets of RAG deployments.

\cleardoublepage
\appendix


\cleardoublepage
\bibliographystyle{plainurl}
\bibliography{reference}

\clearpage
\section*{Appendix}





\section{Proof of Theorems}
\label{appendix:proof}

\paragraph{Notation (issued-query set).}
A retrieval-level partial realization $\psi^R$ records the issued queries and their (retrieval) outcome sets.
We write $\mathrm{dom}(\psi^R)\subseteq Q$ for the set of queries appearing in $\psi^R$ (i.e., issued so far),
and use the shorthand
\[
S(\psi^R)\ :=\ \mathrm{dom}(\psi^R).
\]
In particular, the $S_{t-1}$ in Definition~\ref{def:cmg} is exactly $S(\psi^R_{t-1})$.
We also define the union of revealed retrieved items in $\psi^R$ as
\[
R(\psi^R)\ :=\ \bigcup_{(q,o)\in \psi^R} o .
\]

\noindent\textbf{Lemma A.1 (RAG $\Rightarrow$ independence of unissued outcomes).}
We consider standard RAG deployment: each query is processed independently,
and any internal randomness (e.g., query rewriting sampling or multi-query generation sampling)
is freshly drawn per query and independent across queries.
Formally, let $\Phi=(\xi_q)_{q\in Q}$ where $\{\xi_q\}$ are mutually independent, and assume
$O_{\Phi}(q)$ is determined only by $(q,\xi_q)$.
Then for any retrieval-level partial realization $\psi^R$ and any query $q\notin S(\psi^R)$,
the conditional law of $O_{\Phi}(q)$ is unchanged by conditioning on $\psi^R$.
In particular, for any measurable function $h(\cdot)$,
\begin{equation}
\mathbb{E}\!\left[h\!\left(O_{\Phi}(q)\right)\mid \psi^R\right]
=
\mathbb{E}\!\left[h\!\left(O_{\Phi}(q)\right)\right].
\label{eq:indep-lemma}
\end{equation}

\begin{proof}
Under the stateless assumption, $\psi^R$ depends only on $\{\xi_{q'}: q'\in S(\psi^R)\}$ through
the realized outcomes of already issued queries.
For any $q\notin S(\psi^R)$, $\xi_q$ is independent of $\{\xi_{q'}: q'\in S(\psi^R)\}$,
hence independent of the event defining $\psi^R$.
Therefore conditioning on $\psi^R$ does not change the distribution of $\xi_q$ nor of $O_{\Phi}(q)$,
which yields~\eqref{eq:indep-lemma}.
\end{proof}

\noindent\textbf{Proof of Theorem~\ref{theorem:monotonicity-submodularity}: Adaptive Monotonicity and Submodularity in RAG Crawling}

\begin{proof}
\textbf{(Adaptive monotonicity).}
For any realization $\Phi$ and any query set $A\subseteq Q$, the coverage function satisfies
$f(A\cup\{q\},\Phi)\ge f(A,\Phi)$, since adding a query can only enlarge the union of retrieved items.
By Definition~\ref{def:cmg}, this implies $\Delta(q\mid \psi^{R}_{t-1})\ge 0$ for any retrieval-level
partial realization $\psi^{R}_{t-1}$.

\textbf{(Adaptive submodularity).}
Let $\psi^{R}$ and ${\psi^{R}}'$ be retrieval-level partial realizations such that ${\psi^{R}}'$ is an extension of $\psi^{R}$.
Then $R(\psi^{R})\subseteq R({\psi^{R}}')$, and also $S(\psi^{R})\subseteq S({\psi^{R}}')$.

Fix any query $q\notin S({\psi^{R}}')$ (i.e., $q$ has not been issued under ${\psi^{R}}'$).
Define the shorthand
\[
R \triangleq R(\psi^{R}),\qquad R' \triangleq R({\psi^{R}}'),\qquad Y \triangleq O_{\Phi}(q).
\]
For any realization $\Phi$ consistent with ${\psi^{R}}'$ (hence also consistent with $\psi^{R}$),
the marginal gain of issuing $q$ under $\psi^{R}$ is
\begin{equation}
f\!\left(S(\psi^{R})\cup\{q\},\Phi\right)-f\!\left(S(\psi^{R}),\Phi\right)=|Y\setminus R|,
\label{eq:mg-R}
\end{equation}
and similarly under ${\psi^{R}}'$ it equals $|Y\setminus R'|$.
Since $R\subseteq R'$, we have pointwise diminishing returns for every outcome set $Y$:
\begin{equation}
|Y\setminus R|\ \ge\ |Y\setminus R'|.
\label{eq:pointwise-dr}
\end{equation}

By Definition~\ref{def:cmg} and~\eqref{eq:mg-R},
\begin{equation}
\Delta(q\mid \psi^{R})
=
\mathbb{E}\!\left[\,|Y\setminus R| \mid \psi^{R}\right].
\label{eq:delta-cond}
\end{equation}
Because $q\notin S(\psi^{R})$ and the victim RAG is stateless, Lemma~A.1 applies and yields
\begin{equation}
\Delta(q\mid \psi^{R})
=
\mathbb{E}\!\left[\,|Y\setminus R|\,\right].
\label{eq:delta-uncond}
\end{equation}
Similarly,
\begin{equation}
\Delta(q\mid {\psi^{R}}')
=
\mathbb{E}\!\left[\,|Y\setminus R'|\,\right].
\label{eq:delta-prime-uncond}
\end{equation}

Taking expectation of~\eqref{eq:pointwise-dr} over the (common) distribution of $Y$
gives $\mathbb{E}[|Y\setminus R|]\ge \mathbb{E}[|Y\setminus R'|]$.
Combining with~\eqref{eq:delta-uncond}--\eqref{eq:delta-prime-uncond} yields
\begin{equation}
\Delta(q\mid \psi^{R}) \ge \Delta(q\mid {\psi^{R}}'),
\end{equation}
which is adaptive submodularity.
\end{proof}

\section{Sample Distribution Validation}
\label{appendix:sample-validation}

To ensure that our 1,000-document samples are representative of the full corpora, we compare their semantic distributions using Energy Distance~\cite{szekely2013energy} and the Classifier Two-Sample Test (C2ST)~\cite{lopez2016revisiting}. 
\begin{table}[b]
\centering
\caption{Semantic distribution similarity between each full corpus and its 1,000-document sampled subset, measured by Energy Distance (with permutation test $p$-value) and the Classifier Two-Sample Test (C2ST, using AUC). A small Energy Distance with a high $p$-value and an AUC near 0.5 indicate distributional similarity.}
\label{tab:sample-dist}
\begin{adjustbox}{max width = 0.8\columnwidth}
\begin{tabular}{lcc}
\toprule
\textbf{Corpus} & \textbf{Energy Distance ($value \,/\, p$)} & \textbf{C2ST (AUC)} \\
\midrule
TREC-COVID  & $0.0349 \,/\, 0.5249$ & $0.4993$ \\
SciDocs     & $0.0329 \,/\, 0.5781$ & $0.4938$ \\
NFCorpus    & $0.0281 \,/\, 0.6412$ & $0.5053$ \\
Healthcare  & $0.0276 \,/\, 0.5947$ & $0.4679$ \\
\bottomrule
\end{tabular}
\end{adjustbox}
\end{table}

Energy Distance quantifies the difference between two distributions using Euclidean distances, and is well-suited for high-dimensional or non-Gaussian data. A smaller energy distance with a large permutation test $p$-value suggests no statistically significant difference between the sampled and full sets. As shown in Table~\ref{tab:sample-dist}, all four corpora yield low distances (around 0.03) and high $p$-values ($> 0.5$), indicating strong alignment between samples and full sets.
C2ST (Classifier Two-Sample Test) reframes distribution comparison as a binary classification task . If a classifier cannot distinguish between samples and the full corpus (AUC $\approx 0.5$), the two distributions are considered similar. In Table~\ref{tab:sample-dist}, all AUC scores fall within the 0.46–0.51 range, reinforcing that no meaningful distinction can be learned between the subsets and their corresponding full corpora.


\section{Attacker-side orchestration cost}
\revise{We estimate the attacker-side cost of executing \toolname, i.e., the tokens consumed by the attacker’s model.
As reported in Table~\ref{tab:cost}, \toolname\ processes approximately 6--10 million tokens per dataset.
Based on Doubao-1.5-lite-32K’s API pricing (\$0.042 per million input tokens and \$0.084 per million output tokens), this orchestration cost is \$0.33--\$0.53 per dataset.
As shown in Sec.~\ref{sec:agent}, substituting an open-source model such as Qwen-2.5-7B does not degrade performance, making this component \textbf{virtually zero} aside from GPU hosting.}


\begin{table}[t]
\caption{Token usage and estimated cost per corpus for KG-Constructor (KG-C) and Query Generator (QG) using Doubao-1.5-lite-32k. “InTok” and “OutTok” denote input and output token counts. Each attack costs under \$1.}
\label{tab:cost}
\centering
\begin{adjustbox}{max width = 0.8\columnwidth}
\begin{tabular}{lc@{\hspace{5pt}}c@{\hspace{5pt}}c@{\hspace{5pt}}c}
\toprule
\textbf{Metric} & \textbf{TREC-COVID} & \textbf{SciDocs} & \textbf{NFCorpus} & \textbf{Healthcare} \\
\midrule
KG-C InTok (M)   & 5.561 & 6.209 & 5.997 & 9.094 \\
KG-C OutTok (M)  & 0.724 & 1.277 & 1.076 & 1.373 \\
QG InTok (M)     & 0.847 & 0.596 & 0.366 & 0.783 \\
QG OutTok (M)    & 0.042 & 0.027 & 0.017 & 0.035 \\
\midrule
\textbf{Total Cost (\$)} & \textbf{0.333} & \textbf{0.395} & \textbf{0.359} & \textbf{0.533} \\
\bottomrule
\end{tabular}
\end{adjustbox}

\end{table}



\section{No-topic-prior attacker via one-shot topic probing}
\label{app:no-topic-prior}

\noindent\revise{\textbf{Motivation.}
In Sec.~\ref{subsec:threat-model}, we assume the attacker knows a coarse topic phrase of the target corpus, which is often available from product descriptions, onboarding pages, or light interaction with the assistant~\cite{vertexaisearch_pricing,wang2025silent}.
To be conservative about attacker knowledge, we also evaluate a weaker attacker that starts with no topic prior; the corresponding CR results are reported in Table~\ref{tab:no-topic-prior} (main text).}

\noindent\revise{\textbf{One-shot topic probing.}
The attacker first issues a single benign probe query to the victim RAG system:
\begin{quote}
\small \texttt{What domain do you know about?}
\end{quote}
Let the victim response be $a_{\mathrm{probe}}$.
The attacker then infers a high-level topic phrase from $a_{\mathrm{probe}}$ using an attacker-side LLM and the following prompt.
The output is restricted to a short phrase (3--8 words) to avoid over-specification and keep the prior coarse.
\begin{quote}
\small
\texttt{You are given a sentence that is generated by a RAG system.}\\
\texttt{Infer the high-level domain/topic of the RAG system in 3-8 words.}\\
\texttt{Return ONLY the topic phrase (no quotes, no extra text).}\\
\texttt{Sentence: \{a\_prob\}}\\
\texttt{Topic:}
\end{quote}}

\revise{The inferred phrase is the only topic signal used in the subsequent attack.
For \toolname, it replaces the provided topic input to the Topic-Specific Prior module (Sec.~\ref{sec:design-kg}), which induces the lightweight entity/relation schema and then proceeds with the same closed-loop crawling.
For fair comparison, we provide the same inferred topic phrase to baselines as their input.}

\noindent\revise{\textbf{Budget accounting and metrics.}
We count the probe as one issued query, so the remaining crawling budget is reduced by one under a fixed total budget.
We report Coverage Rate (CR) under the same evaluation protocol as Sec.~\ref{sec:evaluation}.}

\noindent\revise{\textbf{Further analysis: inferred-topic noise and robustness.}
The one-shot probe can yield coarse or off-focus topic phrases that deviate substantially from the manual topic prior. For example, the inferred topic is \emph{Medical Research on Impact Factors} for TREC-COVID (manual topic: \emph{COVID-19}, similarity 0.134) and \emph{Relationship Marketing \& Tech} for SciDocs (manual topic: \emph{Computer Science Research Papers}, similarity 0.185).}

\revise{Despite such noise, Table~\ref{tab:no-topic-prior} shows that RAGCRAWLER remains effective and achieves the best CR across datasets, with only modest changes relative to the topic-prior setting.
In contrast, baselines can be more sensitive to topic quality, consistent with their heavier reliance on the provided topic signal.
This robustness is expected from RAGCRAWLER’s design: the topic prior only bootstraps the initial schema, while the attacker-side global state is continuously updated from retrieved documents, allowing subsequent query planning to self-correct and prioritize unexplored regions of the corpus.}

\section{Additional Experiments Results}
\label{app:additional}

\subsection{Coverage Rate Curves}
\label{appendix:all-coverage}
\begin{figure*}
    \centering
    \includegraphics[width=0.9\linewidth]{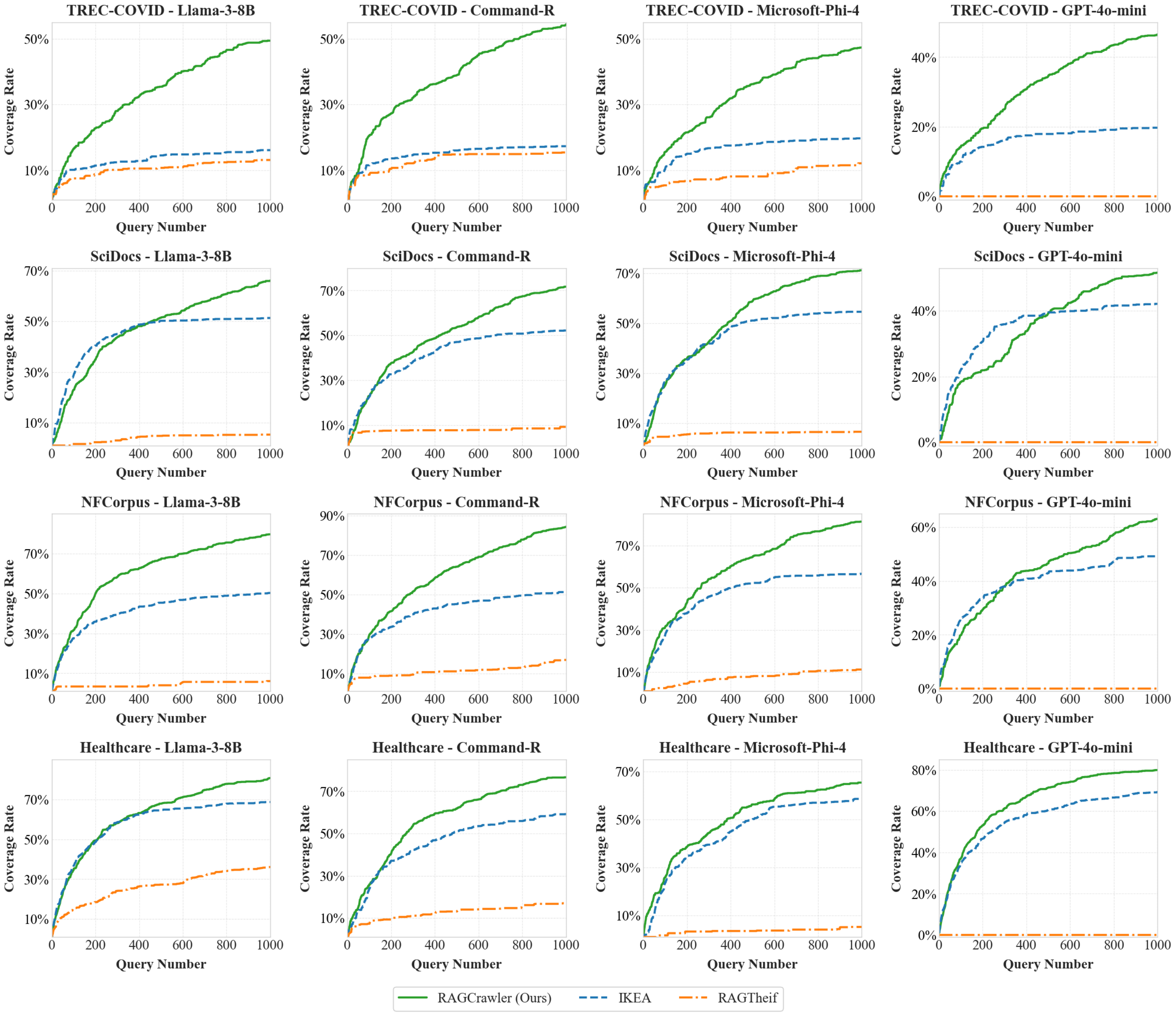}
    \caption{Coverage Rate vs. Query Number (1,000 Budget) across four datasets and four generators.}
    \label{fig:all-coverage}
\end{figure*}

\begin{figure*}
    \centering
    \includegraphics[width=0.9\linewidth]{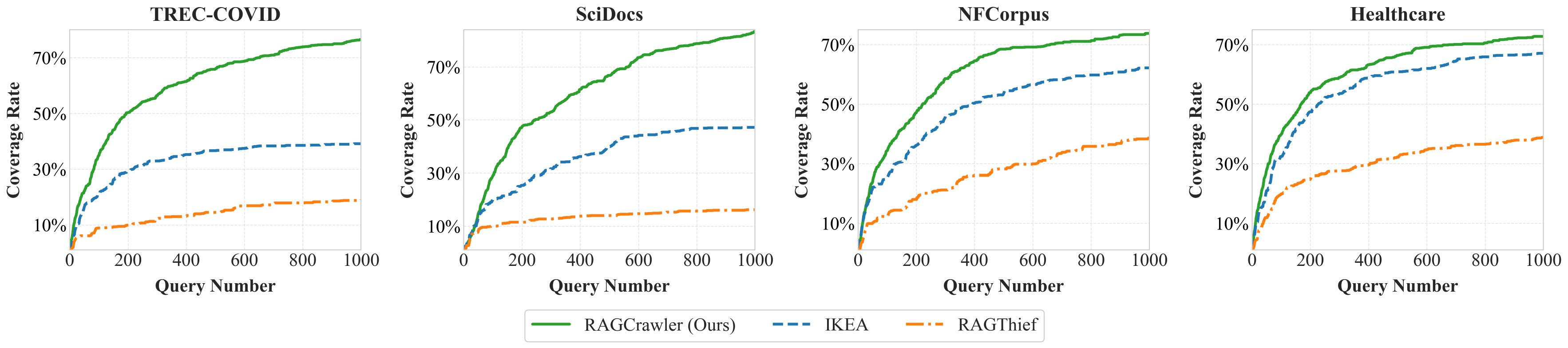}
    \caption{Coverage Rate vs. Query Number (1,000 Budget) across four datasets (GTE Retriever, Llama-3-8B generator).}
    \label{fig:gte-coverage}
\end{figure*}

\begin{figure*}[t]
    \centering
    \includegraphics[width=0.9\linewidth]{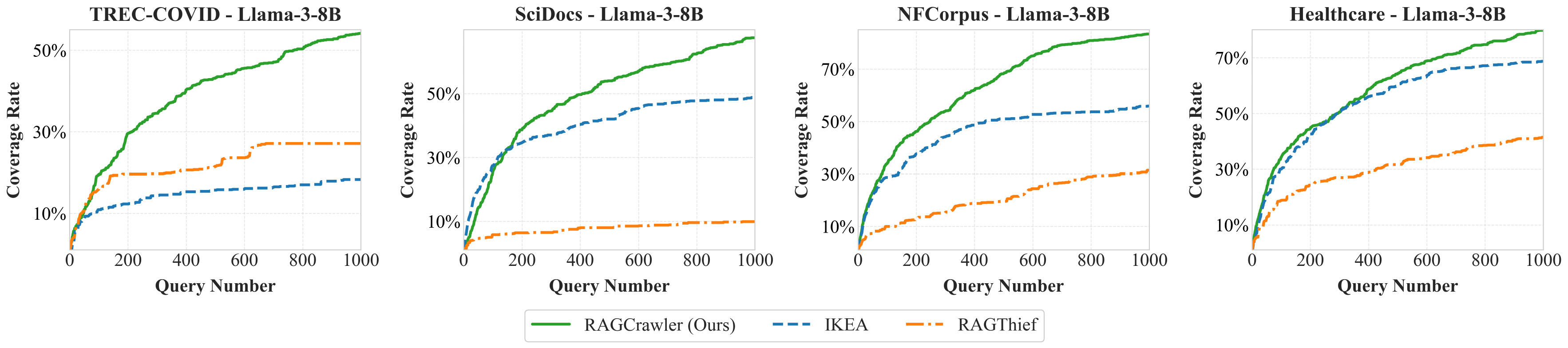}
    \caption{Coverage Rate vs. Query Number (1,000 Budget) across four datasets with Qwen-2.5-7B-Instruct as attacker LLM (BGE Retriever, Llama-3-8B generator).}
    \label{fig:qwen-coverage}
\end{figure*}

For completeness, we provide all Coverage Rate curves across datasets and settings for all experiments in Sec~\ref{sec:effectiveness}, Sec.~\ref{sec:retriever} and Sec.~\ref{sec:agent} in Fig.~\ref{fig:all-coverage}, Fig.~\ref{fig:gte-coverage} and Fig.~\ref{fig:qwen-coverage}. 


\subsection{Additional Ablations}
\label{appendix:hyper-parameter}

\noindent\textbf{Modules.} The effectiveness of \toolname\ stems from the synergistic integration of its three architectural modules.
Without the KG-Constructor, the system loses its global state memory, becoming a stateless, reactive loop unable to distinguish explored regions, a behavior characteristic of IKEA. Without the Strategy Scheduler, the system cannot perform UCB-based prioritization; a fallback to random anchor selection would mimic IKEA’s keyword-based strategy. Finally, without the Query Generator, the scheduler's strategic plan cannot be translated into executable queries, rendering the planning moot.

\begin{figure}[t]
    \centering
    \includegraphics[width=0.85\linewidth]{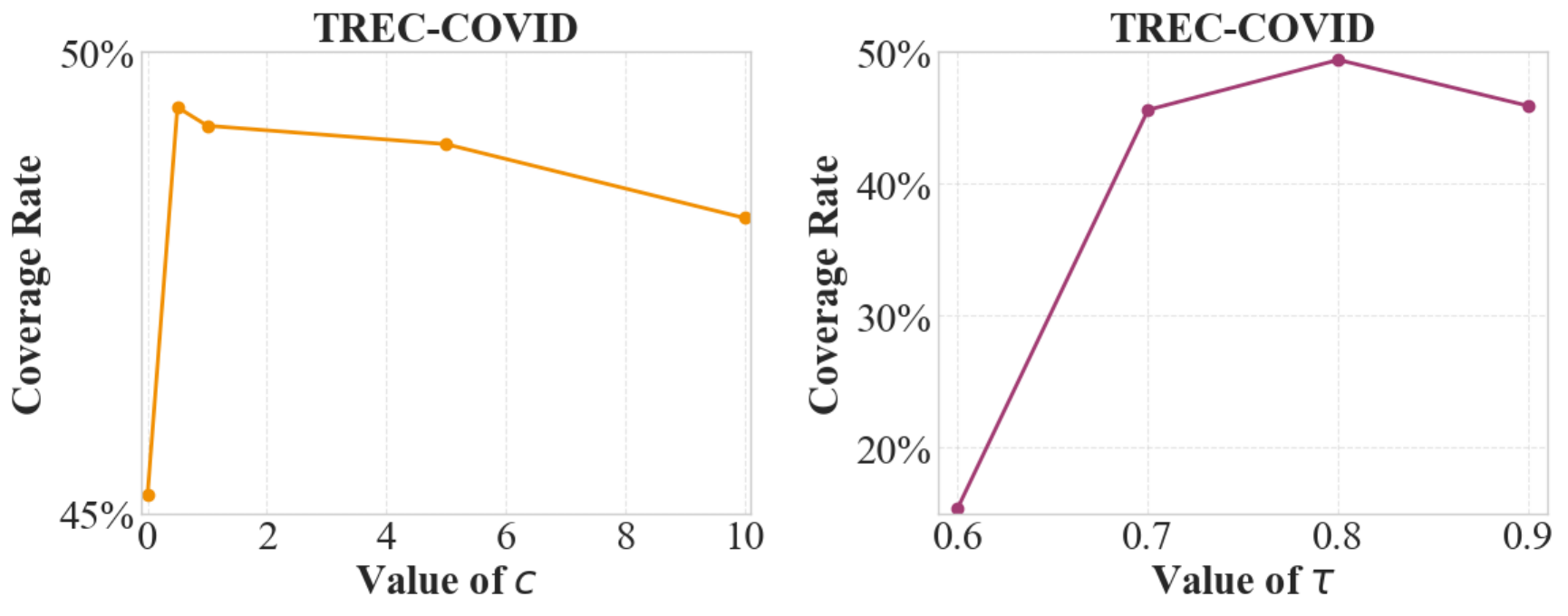}
    \caption{Hyperparameter Choices of \toolname.}
   
    \label{fig:hyperparameter}
\end{figure}

\noindent\textbf{Hyperparameter Choices.}
We examine two key hyperparameters in \toolname: the UCB exploration coefficient $c$ used by the Strategy Scheduler for query selection, and the similarity threshold $\tau_{dup}$ used by the Query Generator to filter high-similarity queries (thereby controlling early stopping).
As shown in Fig.~\ref{fig:hyperparameter}, a moderate UCB coefficient of $c = 0.5$ delivers the best performance by balancing exploitation and exploration. Increasing or decreasing $c$ from this value noticeably degrades performance, underscoring that $c = 0.5$ is near-optimal for a proper exploration–exploitation trade-off. Similarly, the threshold $\tau_{dup}$ exhibits a clear sweet spot. Setting $\tau_{dup}$ too low often causes premature termination. On the other hand, an overly high threshold produces queries that are not sufficiently distinct from one another, which limits exploration efficiency. Our chosen $\tau_{dup}=0.8$ navigates between these extremes.

\subsection{Additional Target-Coverage Query Complexity}
\label{app:qtau}

\revise{
Table~\ref{tab:qtau} summarizes the target-coverage query complexity $Q_{\gamma}$ under a fixed cap $Q_{\max}=5{,}000$ and the resulting final coverage $\mathrm{CR}(Q_{\max})$.
A clear pattern is that baseline heuristics exhibit strong \emph{early} gains but saturate quickly, whereas \toolname continues to accrue coverage as the target increases.
At low targets (50--60\%), IKEA can sometimes approach \toolname on easier corpora (e.g., Healthcare), suggesting that shallow, highly retrievable regions can be harvested with mostly local signals.
However, as $\gamma$ increases, IKEA increasingly plateaus: on three of four datasets it cannot reach 70\% within the cap, and its remaining gains are dominated by redundancy.
RAGThief is consistently the weakest, failing to reach 50\% on most corpora and often stalling far below the cap coverage achieved by the other methods.}

\begin{table}[t]
\centering
\caption{Query complexity for target corpus coverage under a budget cap. For each method and dataset, we report $Q_{\gamma}$, the number of issued queries needed to reach coverage $\gamma\in\{0.5,0.6,0.7,0.8,0.9\}$ with $Q_{\max}=5{,}000$; entries marked $Q_{\gamma}>Q_{\max}$ indicate the target was not reached within the cap. We also report the final coverage $\mathrm{CR}(Q_{\max})$ for all runs. \toolname\ consistently attains the highest $\mathrm{CR}(Q_{\max})$ and reaches substantially higher coverage targets with fewer queries.}

\label{tab:qtau}
\small
\setlength{\tabcolsep}{4pt}
\begin{adjustbox}{max width = \linewidth}
\begin{tabular}{llccccc c}
\toprule
Dataset & Method & $Q_{0.5}$ & $Q_{0.6}$ & $Q_{0.7}$ & $Q_{0.8}$ & $Q_{0.9}$ & $\mathrm{CR}(Q_{\max})$ \\
\midrule
\multirow{3}{*}{TREC-COVID}
& RAGThief  & $>5000$ & $>5000$ & $>5000$ & $>5000$ & $>5000$ & 0.298 \\
& IKEA      & $>5000$ & $>5000$ & $>5000$ & $>5000$ & $>5000$ & 0.196 \\
& \toolname & 1107    & 1972    & 4040    & $>5000$ & $>5000$ & 0.725 \\
\midrule
\multirow{3}{*}{SciDocs}
& RAGThief  & $>5000$ & $>5000$ & $>5000$ & $>5000$ & $>5000$ & 0.072 \\
& IKEA      & 493     & $>5000$ & $>5000$ & $>5000$ & $>5000$ & 0.560 \\
& \toolname & 457     & 776     & 1163    & 1784    & 3330    & 0.936 \\
\midrule
\multirow{3}{*}{NFCorpus}
& RAGThief  & $>5000$ & $>5000$ & $>5000$ & $>5000$ & $>5000$ & 0.192 \\
& IKEA      & 938     & 3427    & $>5000$ & $>5000$ & $>5000$ & 0.640 \\
& \toolname & 200     & 322     & 595     & 1011    & 1753    & 0.977 \\
\midrule
\multirow{3}{*}{Healthcare}
& RAGThief  & 3537    & $>5000$ & $>5000$ & $>5000$ & $>5000$ & 0.530 \\
& IKEA      & 212     & 345     & 1231    & 2925    & $>5000$ & 0.825 \\
& \toolname & 209     & 329     & 568     & 983     & 2434    & 0.941 \\
\bottomrule
\end{tabular}
\end{adjustbox}
\end{table}

\revise{
In contrast, \toolname is the only method that reliably enters the high-coverage regime.
It continues to improve beyond 80\% and even reaches 90\% on three datasets within $Q_{\max}$, while no baseline attains 90\% under the same budget.
This difference is also reflected in final coverage at $Q_{\max}$: \toolname achieves the highest $\mathrm{CR}(Q_{\max})$ on every dataset and maintains a large margin over both baselines, with the gap widening on harder corpora.
TREC-COVID remains the most challenging case: although \toolname cannot reach 80\% within the cap, it still sustains steady growth and substantially outpaces baselines that plateau below 30\%.
Overall, the advantage of global state and planning becomes more pronounced as the desired coverage rises, precisely where local heuristics are most susceptible to redundancy and drift.}

\subsection{Larger victim collections}
\label{app:additional-larger}
\begin{figure}[t]
    \centering
    \includegraphics[width=0.9\linewidth]{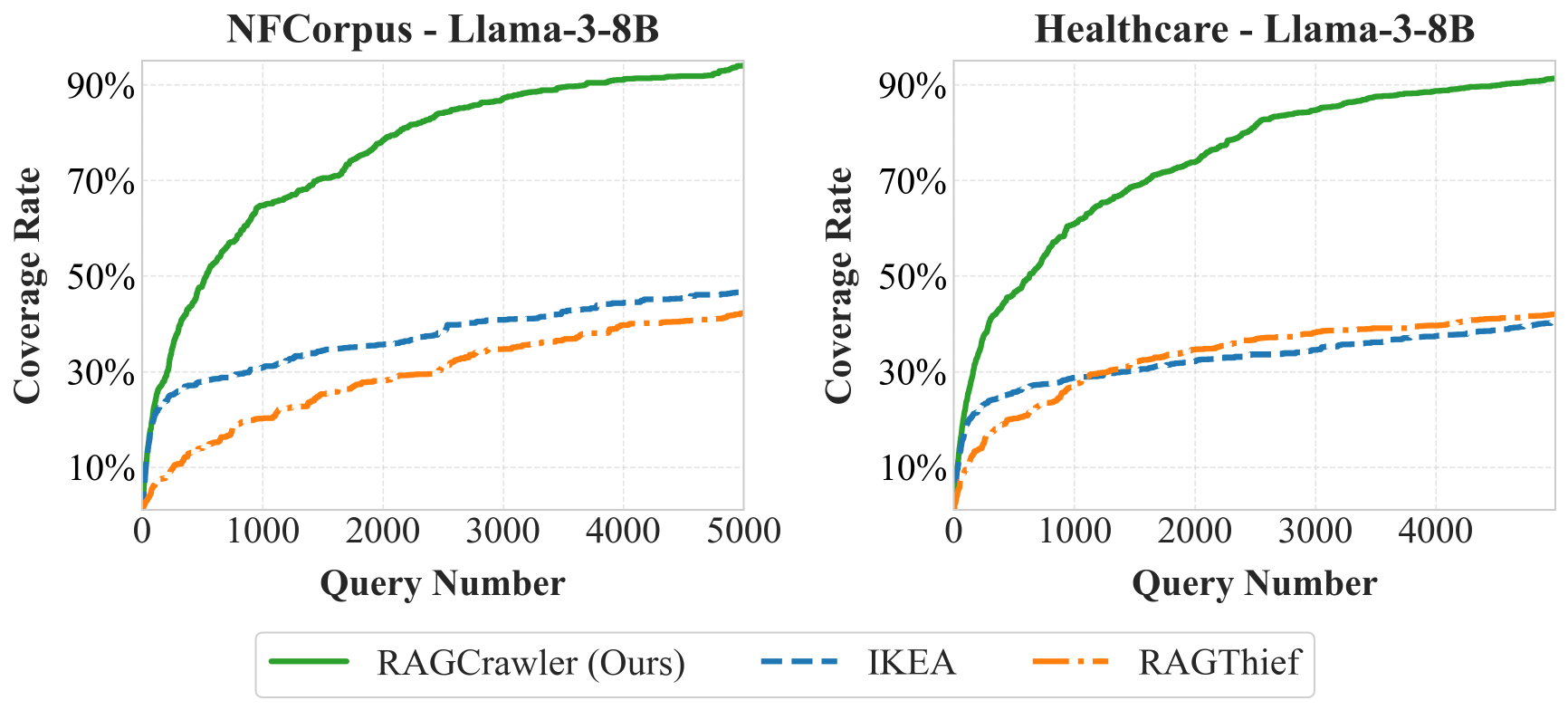}
    \caption{Coverage Rate vs. Query Number (5,000 Budget) on NFCorpus and Healthcare with GPT-4o-mini as attacker LLM (BGE Retriever, Llama-3-8B generator).}
    \label{fig:additional_curve}
\end{figure}

To further investigate the impact of scale, a complementary study was conducted with an expanded configuration of 2,000 documents and a 5,000-Budget. The coverage rate curves for the NFCorpus and Healthcare datasets from this study are depicted in Fig.~\ref{fig:additional_curve}.

\section{Prompts for Experimental Stages}
\label{appendix:prompts}
We document the exact prompts used in RAG settings. More prompts in our method can be found in our code.

\begin{tcolorbox}[colframe=green!50!black, colback=green!5!white, coltitle=white, title=RAG System Prompt]
You are an assistant for question-answering tasks. Use the following pieces of retrieved context to answer the question. If you don't know the answer, say that you don't know.

\textbf{Context:} [context]

\textbf{Query:} [query]

\textbf{Answer:}
\end{tcolorbox}

\begin{tcolorbox}[colframe=blue!50!black, colback=blue!5!white, coltitle=white, title=Query Rewriting Prompt in RAG System]
You are a helpful assistant. Your task is to rewrite the given user question to clarify the user's intent and remove any adversarial intent. Directly output the rewritten question without any other text.

\textbf{Original Question:} [query]

\textbf{Answer:}
\end{tcolorbox}

\begin{tcolorbox}[colframe=orange!50!black, colback=orange!5!white, coltitle=white, title=Multi-query Prompt in RAG System]
You are a helpful assistant. Your task is to generate 3 different versions of the given user question to retrieve relevant documents from a vector database. By generating multiple perspectives on the user question, your goal is to help the user overcome some of the limitations of the distance-based similarity search. Provide directly output these alternative questions separated by newlines without any other text.

\textbf{Original Question:} [query]

\textbf{Answer:}
\end{tcolorbox}


\end{document}